\documentclass{article}

\usepackage{authblk}
\usepackage[a4paper, total={16cm, 24cm}, centering]{geometry}
\usepackage[utf8]{inputenc}
\usepackage[T1]{fontenc}
\usepackage[english,brazil]{babel}
\usepackage{amsmath}
\usepackage{graphicx}
\usepackage{xcolor}
\usepackage{caption}
\usepackage{subcaption}
\usepackage{booktabs}
\usepackage{multirow}
\usepackage{siunitx}
\usepackage[para,online,flushleft]{threeparttable}
\usepackage{cite}
\usepackage{bibunits}
\usepackage{hyperref}
\hypersetup{
    colorlinks=true,
    linkcolor=blue,
    filecolor=magenta,
    urlcolor=cyan,
    citecolor=blue,
    pdftitle={Economic Complexity as a Determinant of Regional Human Development in Brazil},
    pdfauthor={Eduardo Moura Zampirolli and Ruben Interian}
}

\begin{document}

\selectlanguage{english}
\renewcommand{\theHsection}{en.\arabic{section}}
\renewcommand{\theHfigure}{en.\arabic{figure}}
\renewcommand{\theHtable}{en.\arabic{table}}
\begin{bibunit}[plain]
\title{Economic Complexity as a Determinant of Regional Human Development in Brazil:\\
Evidence across Aggregation Scales}

\author{Eduardo Moura Zampirolli \and Ruben Interian}

\affil{University of Campinas (Unicamp)\\
Institute of Computing}

\date{2026}

\maketitle

\begin{abstract}
This study investigates the predictive capacity of models based on Economic Complexity and Complex Network Theory when applied to Brazil's Municipal Human Development Index (\textit{Índice de Desenvolvimento Humano Municipal}, IDHM). For this purpose, the Economic Complexity Index (\textit{Índice de Complexidade Econômica}, ICE) was adapted to the Brazilian context. The main objective is to assess the extent to which different approaches to regional development analysis, such as local productive sophistication and structural integration into the national transportation network, determine socioeconomic development. The methodology compares linear regression models (Ridge, LASSO, and Elastic Net) and nonlinear models (Decision Trees and the \textit{Explainable Boosting Machine}), evaluated through 5-fold cross-validation with grid search. The analyses were conducted at two levels of spatial aggregation: municipalities and Immediate Geographic Regions.

The results show that regionally aggregated models exhibit less statistical noise and greater stability, achieving substantially higher coefficients of determination and greater explanatory power for ICE relative to the other variables. Notably, at the Immediate Region level, ICE alone emerges as the main structural determinant of IDHM, suggesting that regional productive sophistication by itself explains a large share of the variation in human development at this territorial scale. In addition, the inclusion of topological metrics from the road infrastructure network generally improved performance. The \textit{Explainable Boosting Machine} achieved the best predictive performance, reaching $R^2 = 0.8196$ at the Immediate Region level when network metrics were included.
\end{abstract}

\section{Introduction}

To support strategic planning and public policy, it is necessary to understand the factors that are crucial to regional human and economic development. Socioeconomic indices are traditionally used in isolation for this purpose, but they provide a limited view of the causes and structures that drive economic growth. Economic Complexity theory proposes that development is associated with the productive sophistication and technological capabilities of a given location, as represented by the Economic Complexity Index (ICE) \cite{hidalgo2009}.

This study also builds on the results of Morais (2025) \cite{morais2025}, who modeled Brazil's road and waterway infrastructure as a weighted complex network to investigate the relationship between the geographic position of municipalities and their well-being indicators. This makes it possible to test an integrated hypothesis: the IDHM of a locality is determined simultaneously by its internal productive sophistication (ICE) and by its capacity to interact externally with other markets (transportation-network metrics).

\section{Theoretical Background}

This section presents the theoretical foundations for the modeling process and the construction of the predictive variables. The discussion is organized into two parts: Section~\ref{en:sec:complexidade} addresses Economic Complexity and its measures of productive sophistication, whereas Section~\ref{en:sec:rede_viaria_teoria} discusses the modeling of transportation infrastructure from the perspective of Complex Network Theory.

\subsection{Economic Complexity}
\label{en:sec:complexidade}

In the original formulation by Hidalgo and Hausmann \cite{hidalgo2009}, international export data are used to calculate a country's economic complexity index. Applying this metric directly to Brazilian municipalities and regions, however, presents methodological limitations. Brazil's export basket has historically been concentrated in commodities and low-value-added primary products, which do not reflect the country's full productive capacity. Because the national economy is strongly oriented toward the domestic market and much of its technological and structural advancement occurs in the service sector, export volume does not capture the actual complexity of municipalities.

For this reason, this study adopts the adaptation to the Brazilian context validated by Monea \cite{Monea2020}, which uses formal-employment data instead of exports to map local productive knowledge. The primary data source is the Annual Social Information Report (\textit{Relação Anual de Informações Sociais}, RAIS), which provides the number of formal employees by municipality and economic activity according to the National Classification of Economic Activities (\textit{Classificação Nacional de Atividades Econômicas}, CNAE). These data are used to construct the specialization matrix $X_{cp}$, whose elements represent the number of employees in activity $p$ in municipality $c$.

Based on this matrix, Revealed Comparative Advantage (RCA), also known as the Location Quotient (LQ), is calculated to measure the relative specialization of each municipality in each activity:
$$R_{cp} = \frac{X_{cp} \cdot X_{\text{Total}}}{X_c \cdot X_p}$$
where $X_{\text{Total}} = \sum_{c,p} X_{cp}$ is the total number of employees in the matrix, $X_c = \sum_p X_{cp}$ is the total number of employees in municipality $c$, and $X_p = \sum_c X_{cp}$ is the total number of employees in activity $p$. This ratio is then used to obtain the binary specialization matrix $M_{cp}$, defined as
$$M_{cp} = \begin{cases} 1, & \text{if } R_{cp} \geq 1 \\ 0, & \text{otherwise.} \end{cases}$$
A municipality is considered specialized in an activity when its relative share exceeds the average among the other municipalities, that is, when $R_{cp} \geq 1$.

The Economic Complexity Index (ICE) is calculated using the eigenvector method, following Hidalgo and Hausmann \cite{hidalgo2009} and its adaptation to the Brazilian context \cite{Monea2020}. The method begins by constructing a municipality co-occurrence matrix, $\tilde{M}_{cc'}$, which captures the similarity between productive structures based on the activities shared by two municipalities:
$$\tilde{M}_{cc'} = \sum_p \frac{M_{cp} M_{c'p}}{k_{c,0} k_{p,0}}$$
where $k_{c,0} = \sum_p M_{cp}$ is the \textit{diversity} of municipality $c$ (the number of activities in which it is specialized), and $k_{p,0} = \sum_c M_{cp}$ is the \textit{ubiquity} of activity $p$ (the number of municipalities specialized in it). By construction, $\tilde{M}_{cc'}$ is a stochastic matrix, so its dominant eigenvector is constant and uninformative. The complexity vector $K_c$ is the eigenvector associated with the second-largest eigenvalue of $\tilde{M}_{cc'}$, because it captures the main nontrivial dimension of the productive structure.

ICE is the standardized (z-score) version of this eigenvector:
$$\text{ICE}_c = \frac{K_c - \langle K_c \rangle}{\sigma(K_c)}$$

The \textit{Diversity} variable corresponds directly to $k_{c,0}$, the number of activities in which the municipality has revealed comparative advantage.

\subsection{Transportation Infrastructure as a Complex Network}
\label{en:sec:rede_viaria_teoria}

To capture the relational structure of Brazil's territory, the model uses centrality metrics. A network is constructed in which vertices represent municipalities and edges represent transportation links weighted by minimum travel time. The following topological variables are then extracted, following Morais (2025) \cite{morais2025}:
\begin{itemize}
    \item \textbf{Degree Centrality:} Measures the number of direct road and waterway connections of each municipality and reflects its level of local connectivity.
    \item \textbf{Betweenness Centrality:} Measures how frequently a municipality appears on minimum-travel-time paths between pairs of other municipalities and indicates its role as a logistical intermediary.
    \item \textbf{Closeness Centrality:} Computes the inverse of the average distance, measured in travel time, from a municipality to all others. Municipalities with high closeness have greater global structural accessibility.
    \item \textbf{Eigenvector Centrality:} A recursive measure of importance in which a municipality receives a high score when it is connected to other central hubs.
    \item \textbf{Average Neighbor Degree:} Represents the mean number of connections of a municipality's direct neighbors.
    \item \textbf{Clustering Coefficient:} Evaluates connectivity among the neighbors of a municipality, indicating the presence of cohesive local transportation circuits.
    \item \textbf{K-Core:} The $k$-core decomposition identifies the largest subgraph in which every node has at least $k$ connections within the subgraph. A node's \textit{k-core} value is the largest $k$ for which it belongs to such a subgraph, capturing its integration into densely connected regions of the network.
\end{itemize}

Morais showed that these variables are significantly correlated with IDHM, particularly Degree and Closeness Centrality, which supports their inclusion in more complex predictive models.

\section{Objective}

The purpose of this study is to evaluate and compare the performance of explainable machine-learning algorithms for modeling Brazil's IDHM. It seeks to measure the predictive gain obtained by adding transportation-network metrics to traditional models composed exclusively of socioeconomic data, validating the approach both for municipalities and for larger territorial structures (Immediate Regions).

Three specific objectives were established. First, to identify the predictive algorithm with the best generalization capacity and the lowest statistical error. Second, to analyze the performance of ICE and the transportation-network metrics relative to the other predictors, establishing statistical comparisons of the impact of each variable on IDHM. Finally, to determine which territorial scale---municipalities or Immediate Regions---is more appropriate for capturing the dynamics of economic complexity.

\section{Data and Methodology}

This section details the datasets and methods used, ensuring analytical transparency and experimental reproducibility. The process covers the collection of raw data, the construction of the variables, and the training of the predictive models. First, the primary data sources are described. Next, the construction of ICE and the topological transportation-network metrics is presented. The spatial aggregation strategy and the treatment of regional heterogeneity are then explained. Finally, the machine-learning models and validation strategy are described.

\subsection{Data Sources}

Transportation-network data were obtained from the ``Road and Waterway Connections of Brazil'' dataset, published by the Brazilian Institute of Geography and Statistics (IBGE) for 2016 \cite{ibge2016}. The target socioeconomic indicators---IDHM and its Income, Education, and Longevity components---were collected from the Atlas of Human Development in Brazil for 2010 \cite{atlas2013}. Formal-employment data used to calculate ICE were obtained from RAIS for 2020 \cite{rais2020}.

\subsection{Construction of the Economic Complexity Indices}

As described in Section~\ref{en:sec:complexidade}, ICE was calculated for all Brazilian municipalities from the binary specialization matrix $M_{cp}$ derived from RAIS. Although theoretically relevant, the \textit{Diversity} variable ($k_{c,0}$) is highly correlated with ICE. Both are derived from the same matrix $M_{cp}$, and ICE can be interpreted as a measure of sophistication weighted by diversity. For this reason, and to ensure consistency between the two levels of analysis, this variable was excluded from the predictor set.

\subsection{Topological Metrics of the Transportation Network}

Brazil's national infrastructure network was modeled using municipalities as vertices and roads and waterways as edges weighted by travel time, with the \textit{NetworkX} library. For the Immediate Region level, the graph was constructed by aggregating municipal connections: an edge between two regions is created when there is at least one direct link between municipalities belonging to them, and its weight is defined as the sum of the travel times of all such links, representing the total volume of connectivity between the territories.

From the metrics described in Section~\ref{en:sec:rede_viaria_teoria}, the following were selected as final predictive features: \textit{Degree Centrality}, \textit{Betweenness Centrality}, \textit{Closeness Centrality}, \textit{Eigenvector Centrality}, \textit{K-Core}, \textit{Average Neighbor Degree}, and \textit{Clustering Coefficient}.

\subsection{Data Integration and Levels of Spatial Aggregation}

The data were organized at two spatial levels. At the municipality level, the predictors are ICE and the network metrics. At the Immediate Region level, municipal variables were consolidated using population-weighted means based on the 2010 population.

To capture historical structural differences among Brazil's macroregions, including inequalities in industrialization and development policies, binary indicator variables (dummy variables) for the five macroregions were included. With the Central-West as the reference category, \texttt{reg\_Norte}, \texttt{reg\_Nordeste}, \texttt{reg\_Sudeste}, and \texttt{reg\_Sul} were included in every model variant and at both aggregation levels.

\subsection{Predictive Models and Validation Strategy}

The models tested were regularized linear regressions (Ridge, LASSO, and Elastic Net), Decision Trees, and the \textit{Explainable Boosting Machine} (EBM). Missing feature values were imputed using the median of each variable calculated from the training set within each fold. Hyperparameter tuning was performed through grid search with 5-fold cross-validation, using minimization of Mean Absolute Error (MAE) as the primary criterion.

\subsection{Explainable Boosting Machine (EBM)}
\label{en:sec:ebm_metodo}

The \textit{Explainable Boosting Machine} (EBM) is a generalized additive model (GAM) based on gradient boosting, proposed by Lou et al. \cite{lou2012} and implemented in the \textit{InterpretML} library \cite{nori2019interpretml}. Its formulation is
$$\hat{y} = \beta_0 + \sum_i f_i(x_i) + \sum_{i < j} f_{ij}(x_i, x_j)$$
where each $f_i$ is a shape function learned for variable $x_i$, and the $f_{ij}$ terms capture pairwise interactions. Unlike Decision Trees, the EBM remains fully interpretable: the individual contribution of each variable can be visualized separately, which is especially useful for public-policy analysis. The fixed hyperparameters were \textit{inner\_bags} = 5 and \textit{outer\_bags} = 50; the parameter optimized through grid search was the number of pairwise interactions, evaluated over $\{0, 5, 10, 15\}$.

\section{Results and Discussion}

This section presents the results. First, the three families of models are analyzed: linear models, Decision Trees, and EBM. Next, the impact of including the transportation network and the effect of spatial aggregation are evaluated. Finally, the influence of the Northeast regional variable and the relative importance of the predictors are discussed, followed by a comparative synthesis.

\subsection{Regularized Linear Models}

The LASSO, Ridge, and Elastic Net regressors achieved very similar performance at both aggregation levels, as shown in Tables~\ref{en:tab:cv_municipios} and~\ref{en:tab:cv_regioes}. This convergence indicates that the predictor set does not exhibit severe multicollinearity. At the municipality level, the differences among the three models were on the order of $10^{-4}$ for both MAE and $R^2$. At the regional level, Ridge had a slight advantage in $R^2$, with a difference on the order of $10^{-3}$. The similar behavior of the three regressors suggests that the relationships captured are predominantly linear, motivating the use of nonlinear models in the following sections.

\subsection{Decision Trees}

As shown in Table~\ref{en:tab:resultados_arvore}, the Decision Tree for Immediate Regions in the With Network variant achieved its best result with \textit{max\_depth} = 5 and \textit{min\_leaf} = 10 (MAE = 0.0247). At the municipality level, the optimal depth was 7, reflecting the greater heterogeneity of the samples at this scale.

The impurity-reduction plots in Figure~\ref{en:fig:red_impur_arvores} reveal different patterns at the two levels. At the municipality level, \texttt{reg\_Nordeste} surpasses ICE as the main predictor, with importance above 0.50 in both the With Network and Without Network variants. At the Immediate Region level, ICE dominates, with values above 0.90, whereas \texttt{reg\_Nordeste} loses relevance. The transportation-network metrics have low individual importance, but their inclusion produces consistent gains in $R^2$, especially at the municipality level.

\subsection{Explainable Boosting Machine (EBM)}

The EBM achieved the best performance among all tested models. For Immediate Regions in the With Network variant, the model reached MAE = 0.0226 and $R^2 = 0.8196$---the best values in the study, as shown in Table~\ref{en:tab:ebm_resultados}. Its improvement over the linear models results from the EBM's ability to capture nonlinear relationships and interactions among variables automatically. Configurations with 5 to 15 interactions provided the best balance between predictive power and generalization.

\begin{figure}[!htbp]
    \centering
    \begin{subfigure}[b]{0.48\textwidth}
        \includegraphics[width=\textwidth]{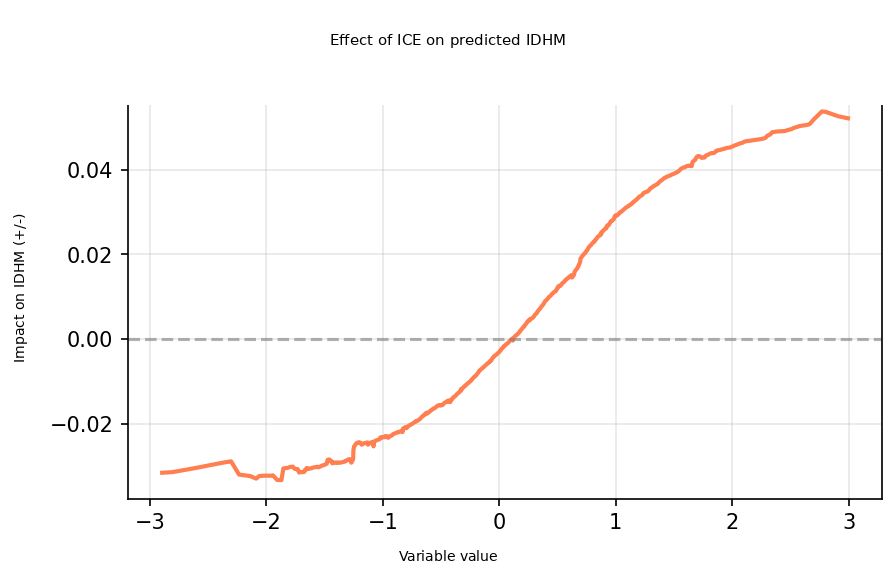}
        \caption{Municipalities with the transportation network}
        \label{en:fig:shape_mun_rede}
    \end{subfigure}
    \hfill
    \begin{subfigure}[b]{0.48\textwidth}
        \includegraphics[width=\textwidth]{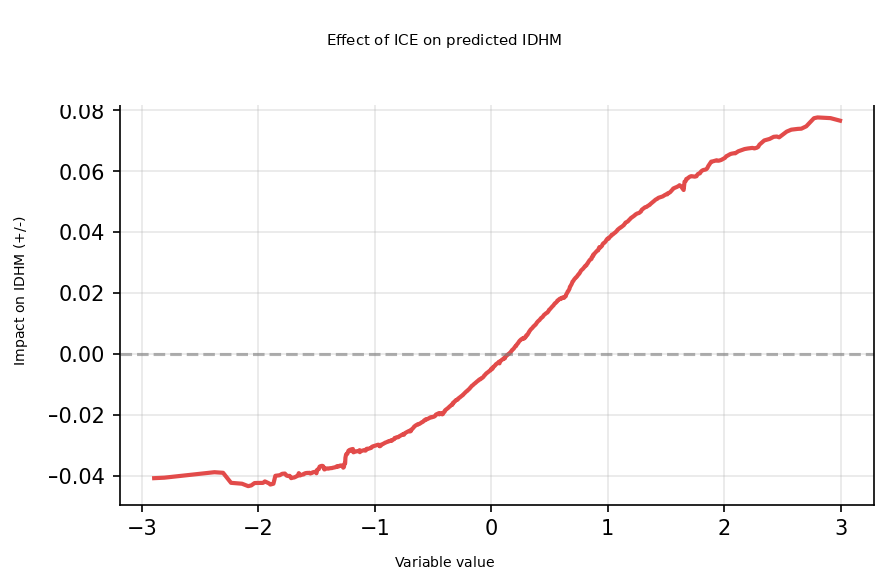}
        \caption{Municipalities without the transportation network}
        \label{en:fig:shape_mun_sem}
    \end{subfigure}

    \vspace{1em}

    \begin{subfigure}[b]{0.48\textwidth}
        \includegraphics[width=\textwidth]{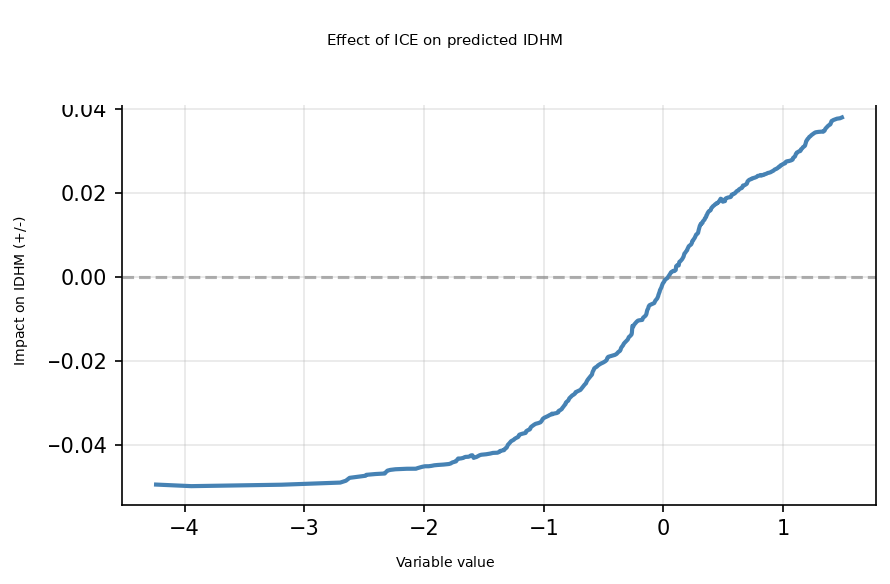}
        \caption{Immediate Regions with the transportation network}
        \label{en:fig:shape_imm_rede}
    \end{subfigure}
    \hfill
    \begin{subfigure}[b]{0.48\textwidth}
        \includegraphics[width=\textwidth]{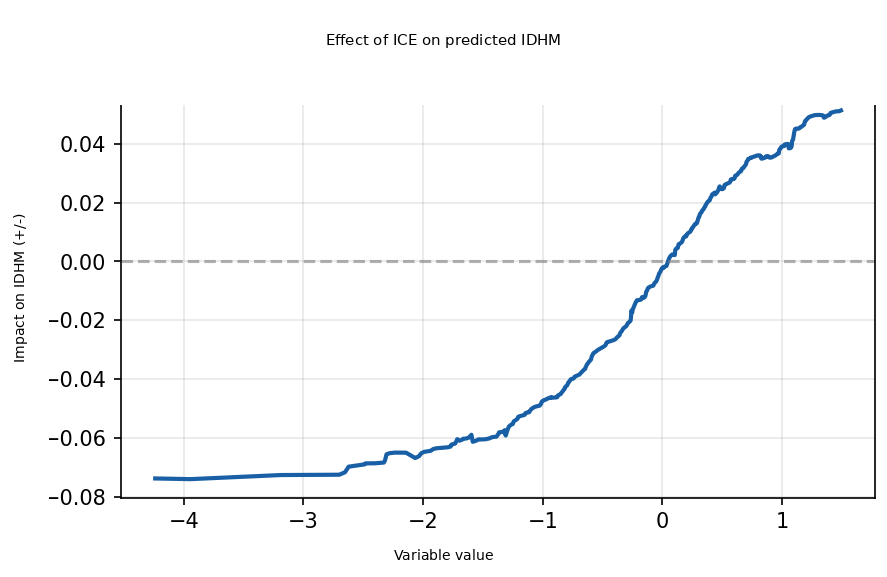}
        \caption{Immediate Regions without the transportation network}
        \label{en:fig:shape_imm_sem}
    \end{subfigure}

    \caption{ICE shape functions in the EBM by aggregation level and model variant. The vertical axis represents the additive impact on predicted IDHM, and the horizontal axis represents the ICE value.}
    \label{en:fig:shape_ice}
\end{figure}

The ICE shape functions in Figure~\ref{en:fig:shape_ice} confirm this nonlinearity: the relationship with IDHM is volatile and has a low slope for very negative ICE values, increases almost linearly in the central range ($-1$ to $1$), and exhibits diminishing returns above $1$. At the Immediate Region level, the same pattern appears with less variability at the extremes, reflecting the smoothing effect of spatial aggregation. In all configurations, the curve has a smaller amplitude in the With Network variant, indicating that the network metrics absorb part of the IDHM variation previously attributed exclusively to ICE.

Regarding the relative importance of variables (Figure~\ref{en:fig:red_impur_ebm}), the Decision Tree pattern is repeated: at the municipality level, \texttt{reg\_Nordeste} remains the most influential predictor, whereas ICE leads at the regional level. The difference between the two variables is smaller in the EBM than in the Decision Trees, indicating that the additive model distributes explanatory power more evenly and makes better use of interactions among ICE, network metrics, and regional indicators.

\begin{table}[htbp]
\centering
\caption{EBM Performance Comparison (Best Hyperparameters)}
\label{en:tab:ebm_resultados}
\begin{tabular}{llccc}
\toprule
\textbf{Level} & \textbf{Variant} & \textbf{Optimized Interactions} & \textbf{MAE} & \textbf{$R^2$} \\
\midrule
Municipality & Without Network & 10 & 0.0289 & 0.7363 \\
Municipality & With Network & 15 & 0.0258 & 0.7890 \\
Immediate Region & Without Network & 5 & 0.0238 & 0.8050 \\
Immediate Region & With Network & 15 & 0.0226 & 0.8196 \\
\bottomrule
\end{tabular}
\end{table}

\subsection{Inclusion of Network Metrics in the Analysis}

As observed in the preceding sections, including network metrics systematically improves performance at both levels and across all model families. At the municipality level, the With Network variant increased the $R^2$ of the linear regressors from approximately 0.726 to 0.750. At the Immediate Region level, the effect was even more pronounced: the best linear model increased from $R^2 = 0.789$ to $R^2 = 0.811$, as shown in Table~\ref{en:tab:melhores_geral}. This increasing effect at higher aggregation levels indicates that transportation-network centrality metrics capture dimensions of development that are particularly relevant in a regional context, in which market access and logistical position collectively affect groups of municipalities.

\subsection{Effect of Spatial Aggregation and Productive Interdependence}

Comparing Tables~\ref{en:tab:cv_municipios} and~\ref{en:tab:cv_regioes} reveals a clear difference in performance between the municipal and regional scales. For the linear models, $R^2$ increases from approximately 0.73 to 0.79 in the Without Network variant and from 0.75 to 0.81 in the With Network variant---a gain of 6 to 7 percentage points solely from changing the territorial scale. By comparison, including transportation-network metrics within each scale produces a gain of only 2 to 3 percentage points. The EBM follows the same pattern, increasing from $R^2 = 0.789$ at the municipality level with the network to $R^2 = 0.820$ at the regional level with the network. Thus, the choice of aggregation scale affects predictive performance more strongly than any other methodological factor tested in this study.

This result is not only quantitative: the change of scale also qualitatively alters the role of ICE in the models. The importance diagrams for the Decision Trees and EBM (Figures~\ref{en:fig:red_impur_arvores} and~\ref{en:fig:red_impur_ebm}) show that, at the municipality level, ICE shares the role of main predictor with \texttt{reg\_Nordeste}, whereas at the Immediate Region level ICE alone accounts for more than 90\% of impurity reduction. In other words, economic complexity becomes sufficient to explain variation in IDHM when the data are regionally aggregated, but not at the local level. This points to a structural limitation of municipal modeling: at this scale, the ICE of an isolated municipality contains insufficient information about its actual productive environment.

The reason lies in the operation of regional economies. A municipality's IDHM depends not only on its internal productive structure but is also strongly shaped by the complexity of surrounding territories through access to more sophisticated labor markets, flows of capital and specialized services, and the diffusion of technical knowledge among nearby firms. When development is modeled at the municipal scale, the algorithm has no direct access to these externalities, resulting in greater residual error and lower explanatory power for ICE. When municipalities are aggregated into Immediate Regions, these interdependencies become implicitly represented in the regional mean ICE itself, which reflects the sophistication of the productive ecosystem as a whole rather than that of each isolated unit.

\begin{figure}[!htbp]
    \centering
    \includegraphics[width=0.98\linewidth]{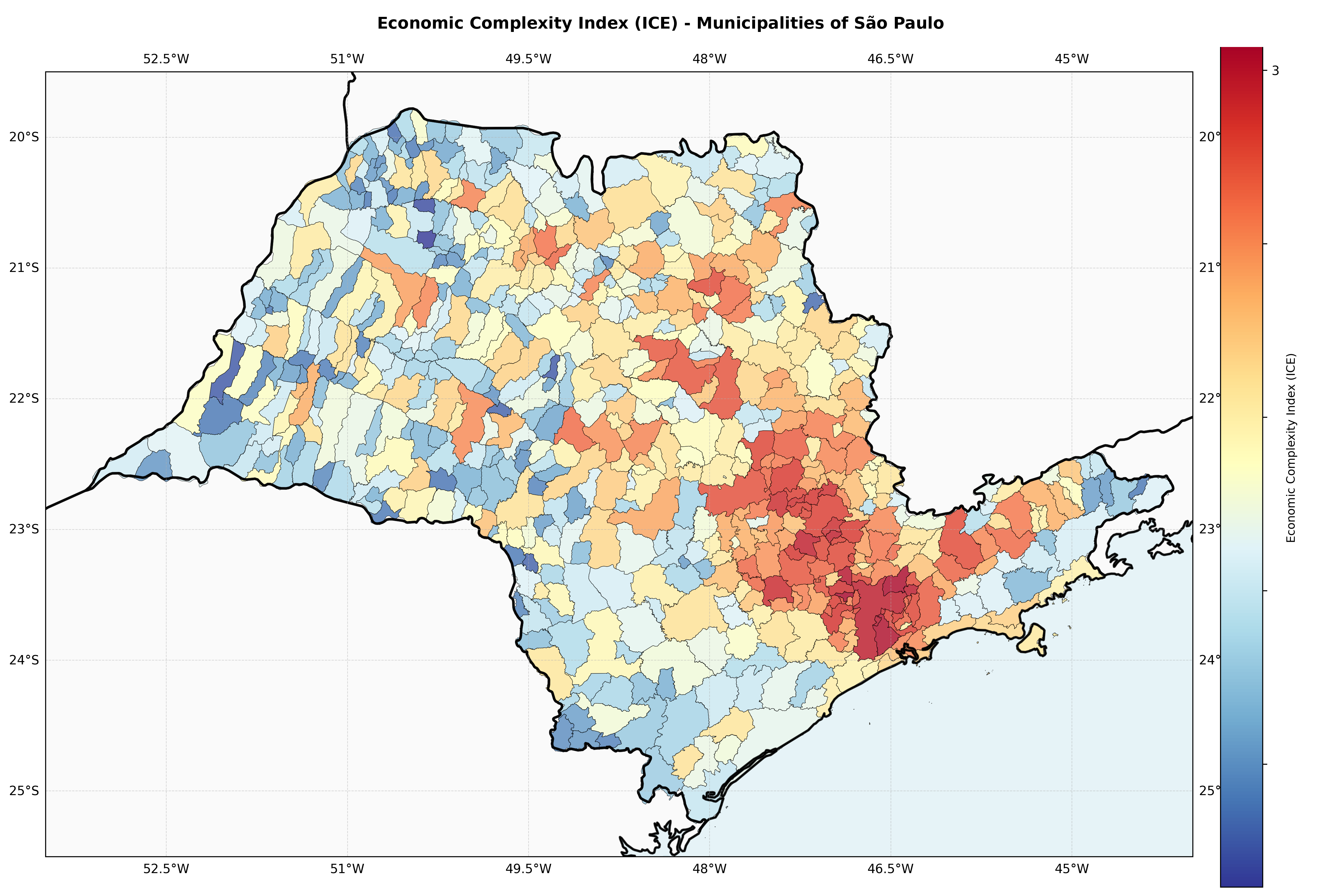}
    \caption{Economic complexity of the municipalities in the state of São Paulo.}
    \label{en:fig:mapa_sp}
\end{figure}

This dynamic is illustrated by the São Paulo Metropolitan Region, detailed in Table~\ref{en:tab:ice_idhm_municipios} and shown in Figure~\ref{en:fig:mapa_sp}. The municipality of São Paulo has an IDHM of 0.805, but its ICE (2.949) is surpassed by surrounding industrial municipalities: Guarulhos (ICE = 3.127, IDHM = 0.763), São Bernardo do Campo (ICE = 3.076, IDHM = 0.805), and Diadema (ICE = 3.024, IDHM = 0.757). These municipalities concentrate the region's more technologically sophisticated production but have IDHM values equal to or lower than that of the capital, which specializes in services, finance, and management. At the municipality level, this pattern creates a disconnect between ICE and IDHM that the model cannot explain locally: the municipality with the highest ICE does not necessarily have the highest IDHM. When these municipalities are aggregated into a single Immediate Region, however, the regional mean ICE reflects an integrated productive ecosystem, and its relationship with the region's population-weighted mean IDHM becomes much more stable and interpretable. This statistical coherence---absent at the municipal scale and present at the regional scale---explains the observed improvement in model performance.

\begin{table}[htbp]
\centering
\caption{Economic Complexity Index (ICE) and IDHM by Municipality}
\label{en:tab:ice_idhm_municipios}
\begin{tabular}{llcc}
\toprule
\textbf{Municipality} & \textbf{Municipality ID} & \textbf{ICE} & \textbf{IDHM} \\
\midrule
São Paulo & 3550308 & 2.949147 & 0.805 \\
\midrule
Guarulhos & 3518800 & 3.127885 & 0.763 \\
São Bernardo do Campo & 3548708 & 3.076483 & 0.805 \\
Diadema & 3513801 & 3.024338 & 0.757 \\
São Caetano do Sul & 3548807 & 2.840713 & 0.862 \\
Cotia & 3513009 & 2.721974 & 0.780 \\
Mauá & 3529401 & 2.705295 & 0.766 \\
Santo André & 3547809 & 2.643223 & 0.815 \\
Taboão da Serra & 3552809 & 2.561624 & 0.769 \\
Poá & 3539806 & 2.507171 & 0.776 \\
Osasco & 3534401 & 2.485954 & 0.776 \\
Ferraz de Vasconcelos & 3515707 & 2.329641 & 0.738 \\
Juquitiba & 3525300 & 2.052169 & 0.709 \\
Mairiporã & 3528502 & 2.030977 & 0.788 \\
Itapecerica da Serra & 3522208 & 1.751917 & 0.742 \\
Embu-Guaçu & 3515103 & 1.566397 & 0.749 \\
Itaquaquecetuba & 3522406 & 0.837479 & 0.714 \\
\bottomrule
\end{tabular}
\end{table}

\begin{table}[htbp]
\centering
\caption{Model Performance at the Municipality Level (5-Fold CV)}
\label{en:tab:cv_municipios}
\begin{tabular}{llcc}
\toprule
\textbf{Variant} & \textbf{Model} & \textbf{MAE} & \textbf{$R^2$} \\
\midrule
\multirow{3}{*}{With Network}
 & LASSO      & 0.0281 & 0.7498 \\
 & Elastic Net & 0.0281 & 0.7498 \\
 & Ridge      & 0.0282 & 0.7490 \\
\midrule
\multirow{3}{*}{Without Network}
 & Ridge      & 0.0294 & 0.7256 \\
 & LASSO      & 0.0294 & 0.7255 \\
 & Elastic Net & 0.0295 & 0.7254 \\
\bottomrule
\end{tabular}
\end{table}

\begin{table}[htbp]
\centering
\caption{Model Performance at the Immediate Region Level (5-Fold CV)}
\label{en:tab:cv_regioes}
\begin{tabular}{llcc}
\toprule
\textbf{Variant} & \textbf{Model} & \textbf{MAE} & \textbf{$R^2$} \\
\midrule
\multirow{3}{*}{With Network}
 & Ridge      & 0.0237 & 0.8112 \\
 & LASSO      & 0.0237 & 0.8087 \\
 & Elastic Net & 0.0237 & 0.8088 \\
\midrule
\multirow{3}{*}{Without Network}
 & Ridge      & 0.0250 & 0.7893 \\
 & Elastic Net & 0.0251 & 0.7874 \\
 & LASSO      & 0.0251 & 0.7873 \\
\bottomrule
\end{tabular}
\end{table}

\begin{table}[htbp]
\centering
\caption{Decision Tree Optimization Results by Level and Variant}
\label{en:tab:resultados_arvore}
\begin{tabular}{lccccc}
\toprule
\textbf{Level} & \textbf{Variant} & \textbf{max\_depth} & \textbf{min\_leaf} & \textbf{MAE} & \textbf{$R^2$} \\
\midrule
Municipality & Without Network & 5 & 2  & 0.0294 & 0.7277 \\
Municipality & With Network & 7 & 10 & 0.0270 & 0.7647 \\
\midrule
Immediate Region & Without Network & 7 & 10 & 0.0254 & 0.7646 \\
Immediate Region & With Network & 5 & 10 & 0.0247 & 0.7788 \\
\bottomrule
\end{tabular}
\end{table}

\subsection{Regional Influence}

One of the most important findings is the weight of the \texttt{reg\_Nordeste} variable in the models, especially at the municipality level. In the Decision Trees (Figure~\ref{en:fig:red_impur_arvores}), this variable surpasses ICE as the main predictor of municipal IDHM, with impurity-reduction importance above 0.50 in both variants. In the EBM, this predominance remains, although with a smaller magnitude.

This result reflects structural characteristics of Brazilian society: municipalities in the Northeast have IDHM values systematically below the national average for reasons that extend beyond local productive sophistication. The region has historically lower industrialization, less extensive infrastructure coverage, greater dependence on government transfers, and a predominance of extensive agricultural activities. In other words, being located in the Northeast adds predictive information about IDHM that ICE alone does not capture at the municipality level.

This reveals a limitation of the local model: at the municipal scale, the algorithm tends first to separate ``Northeastern municipalities'' from the others and only then use productive sophistication as a refinement. The first split nodes of the Decision Trees operate essentially as a regional classifier before incorporating ICE and network metrics.

At the Immediate Region level, this effect decreases substantially: ICE becomes the dominant predictor and \texttt{reg\_Nordeste} loses relative importance. Spatial aggregation reduces part of the region's internal heterogeneity, and the structural differences of the Northeast become captured by variation in ICE itself across Immediate Regions, which is consistently lower in this macroregion. This reinforces the conclusion that the Immediate Region scale is more appropriate for analyzing the relationship between productive complexity and human development in an integrated manner.

\subsection{Variable Importance and Interpretability}

\begin{figure}[!htbp]
    \centering
    \begin{subfigure}[b]{0.48\textwidth}
        \includegraphics[width=\textwidth]{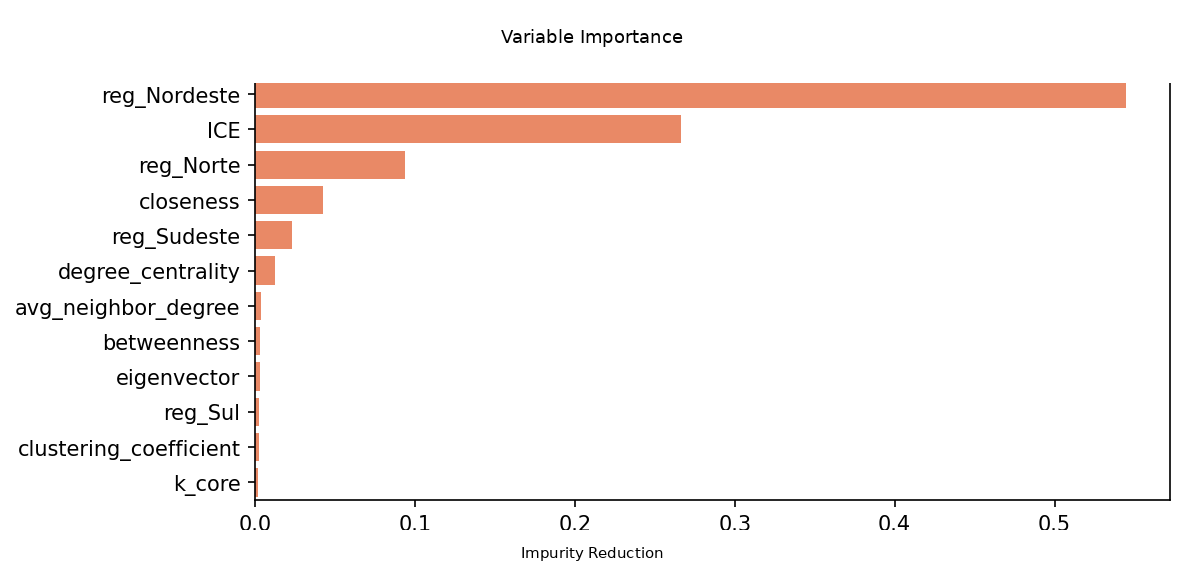}
        \caption{Municipalities with the transportation network}
        \label{en:fig:imp_mun_tree_rede}
    \end{subfigure}
    \hfill
    \begin{subfigure}[b]{0.48\textwidth}
        \includegraphics[width=\textwidth]{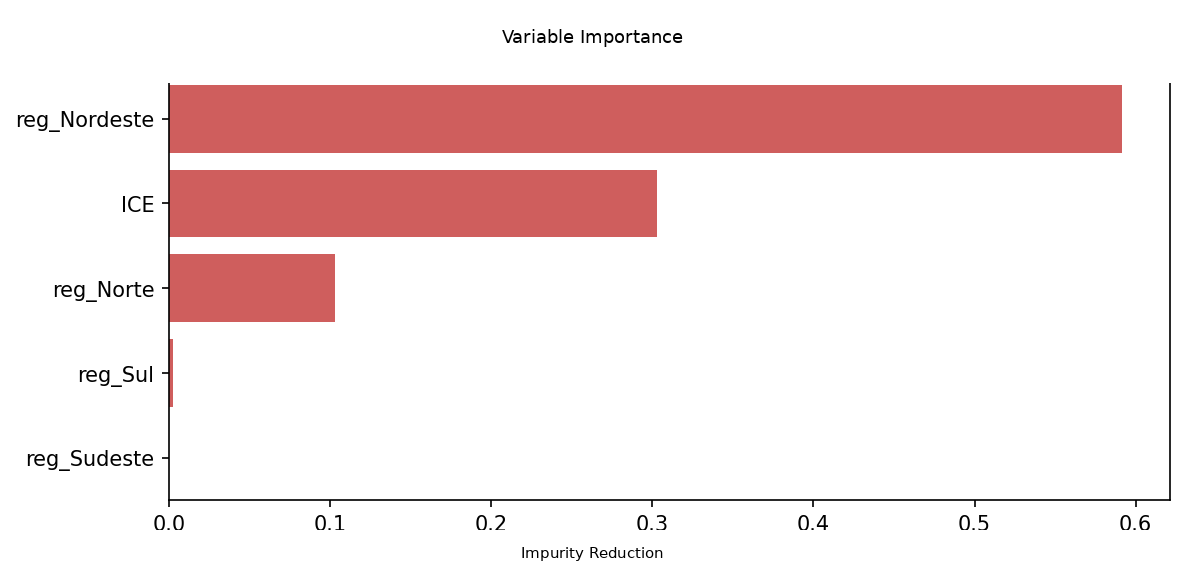}
        \caption{Municipalities without the transportation network}
        \label{en:fig:imp_mun_tree_sem}
    \end{subfigure}

    \vspace{1em}

    \begin{subfigure}[b]{0.48\textwidth}
        \includegraphics[width=\textwidth]{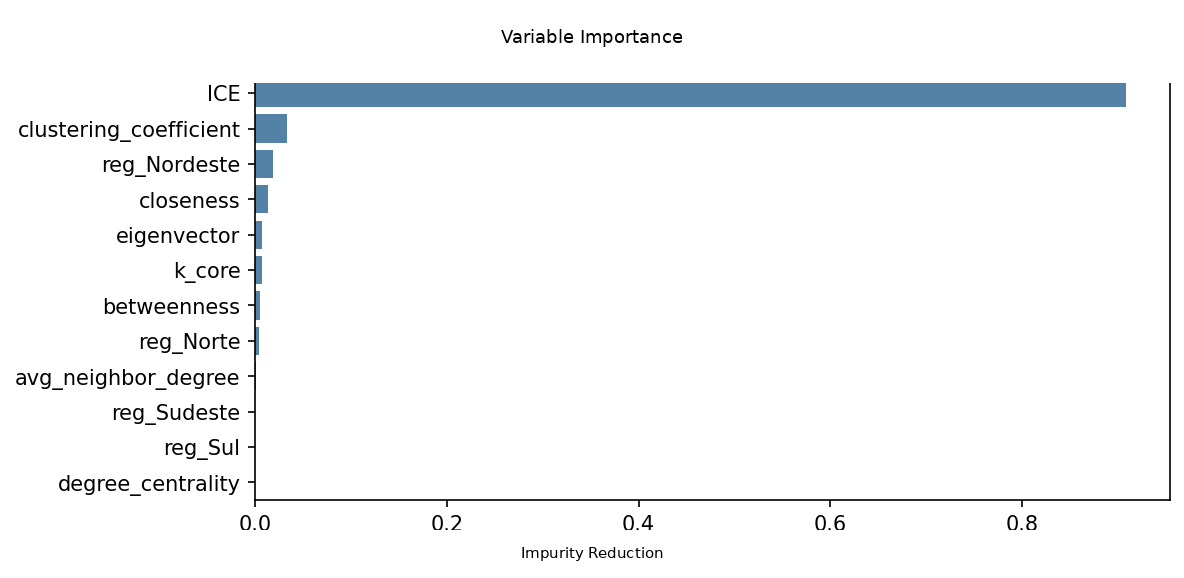}
        \caption{Immediate Regions with the transportation network}
        \label{en:fig:imp_imm_tree_rede}
    \end{subfigure}
    \hfill
    \begin{subfigure}[b]{0.48\textwidth}
        \includegraphics[width=\textwidth]{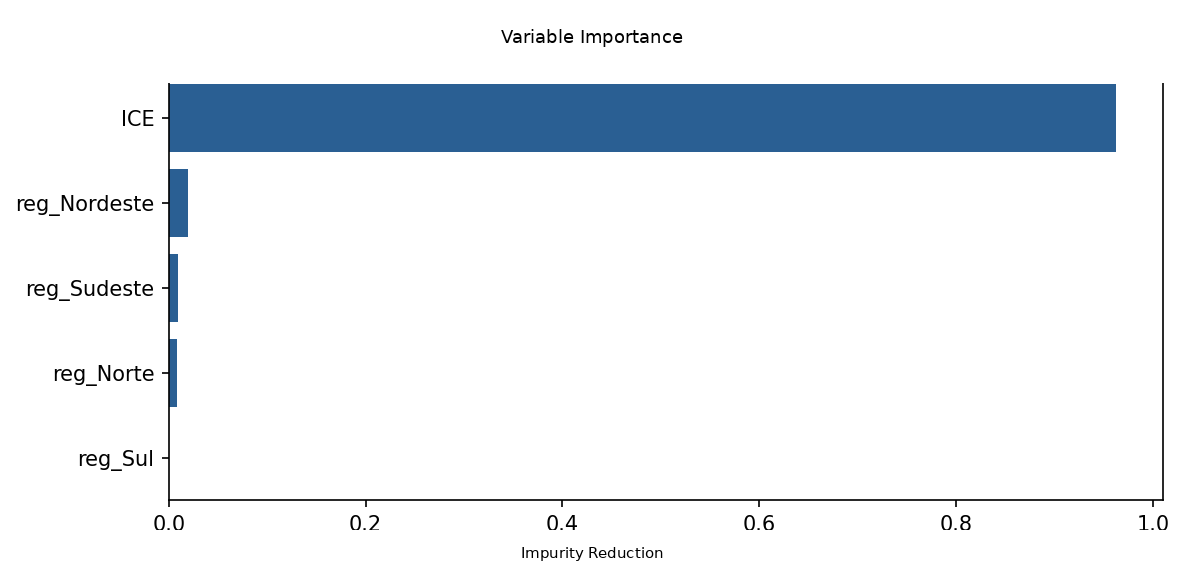}
        \caption{Immediate Regions without the transportation network}
        \label{en:fig:imp_imm_tree_sem}
    \end{subfigure}

    \caption{Impurity reduction attributed to the predictive variables in the Decision Trees, by aggregation level and model variant.}
    \label{en:fig:red_impur_arvores}
\end{figure}

\begin{figure}[!htbp]
    \centering
    \begin{subfigure}[b]{0.48\textwidth}
        \includegraphics[width=\textwidth]{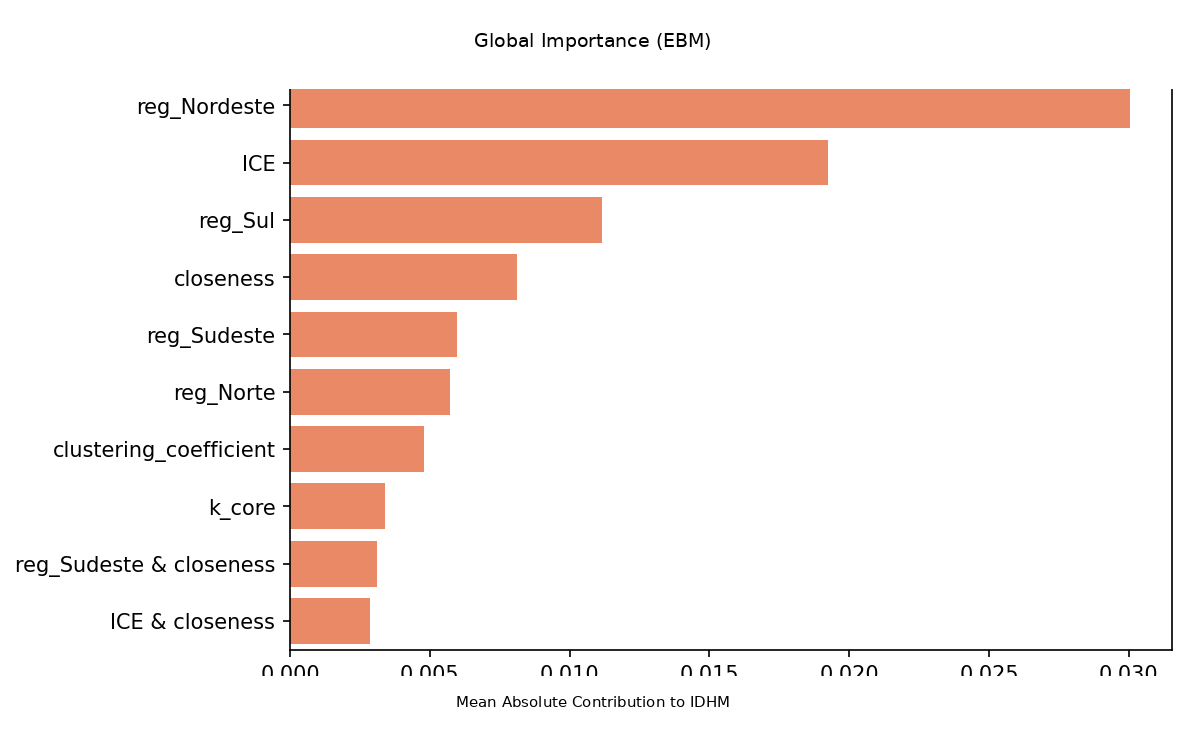}
        \caption{Municipalities with the transportation network}
        \label{en:fig:imp_mun_ebm_rede}
    \end{subfigure}
    \hfill
    \begin{subfigure}[b]{0.48\textwidth}
        \includegraphics[width=\textwidth]{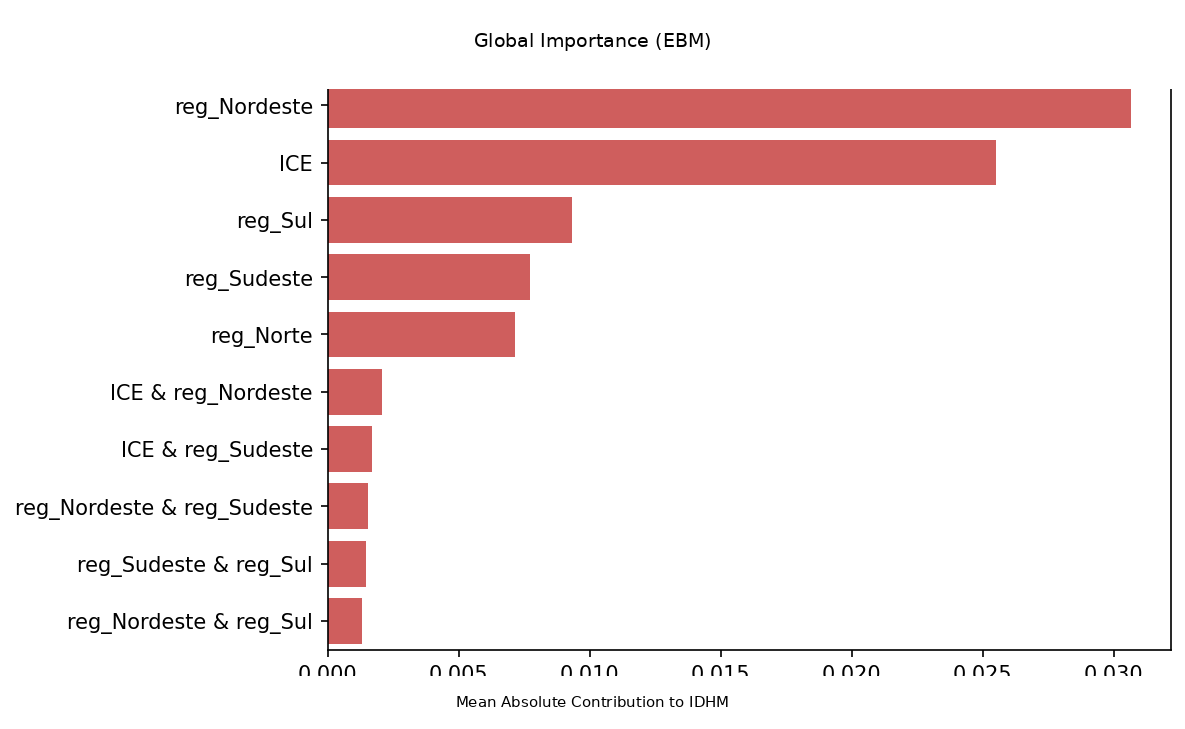}
        \caption{Municipalities without the transportation network}
        \label{en:fig:imp_mun_ebm_sem}
    \end{subfigure}

    \vspace{1em}

    \begin{subfigure}[b]{0.48\textwidth}
        \includegraphics[width=\textwidth]{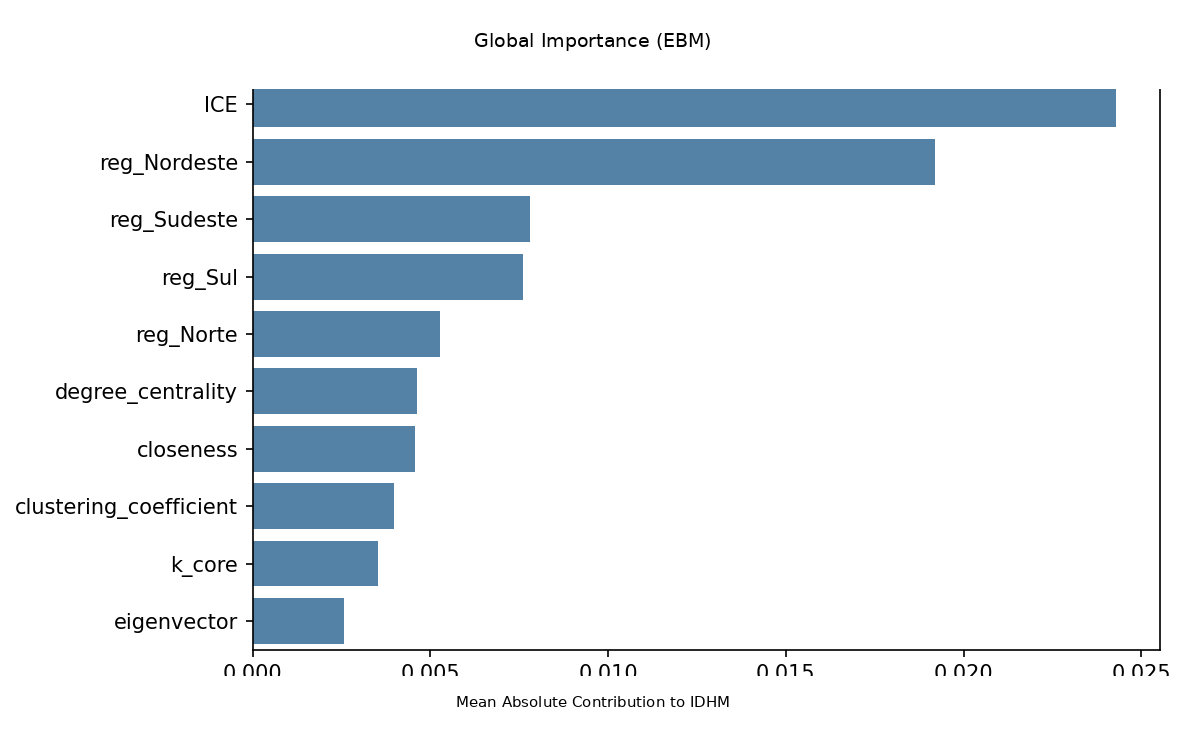}
        \caption{Immediate Regions with the transportation network}
        \label{en:fig:imp_imm_ebm_rede}
    \end{subfigure}
    \hfill
    \begin{subfigure}[b]{0.48\textwidth}
        \includegraphics[width=\textwidth]{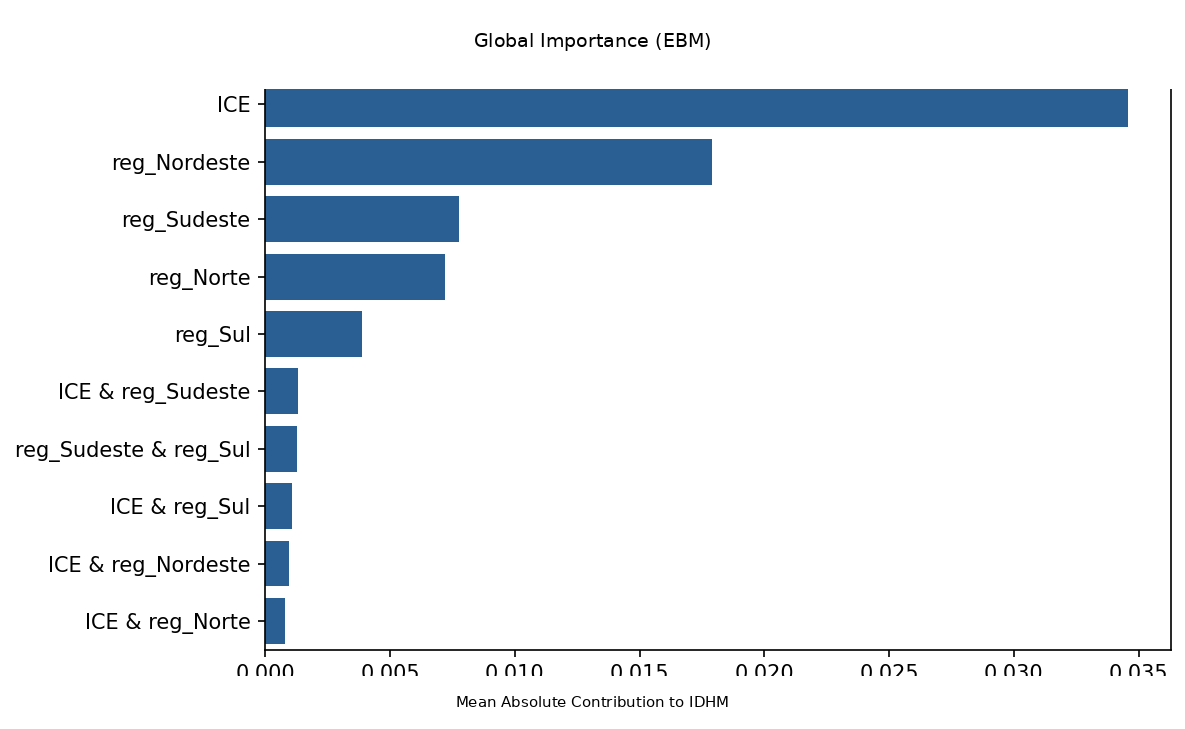}
        \caption{Immediate Regions without the transportation network}
        \label{en:fig:imp_imm_ebm_sem}
    \end{subfigure}

    \caption{Importance of the predictive variables in the EBM, by aggregation level and model variant.}
    \label{en:fig:red_impur_ebm}
\end{figure}

The impurity-reduction diagrams for the Decision Trees (Figure~\ref{en:fig:red_impur_arvores}) and the EBM importance diagrams (Figure~\ref{en:fig:red_impur_ebm}) show a consistent pattern: at the municipality level, \texttt{reg\_Nordeste} is the most relevant predictor, followed by ICE; at the Immediate Region level, ICE leads by a wide margin. Among the network metrics, \textit{Closeness Centrality} and \textit{Degree Centrality} were the most consistent across both aggregation levels and model families, although their individual importance was lower than that of the socioeconomic variables.

\subsection{Comparative Synthesis of the Models}

Table~\ref{en:tab:melhores_geral} summarizes the best result from each model family for each configuration. Two patterns stand out. First, modeling at the Immediate Region level consistently outperforms municipal modeling in every combination: the gain in $R^2$ from changing scale is greater than the gain associated with any other factor analyzed. Second, EBM is the best model in every configuration, with a particularly large margin for municipalities in the With Network variant ($R^2 = 0.789$), compared with the linear model ($R^2 = 0.750$) and the Decision Tree ($R^2 = 0.765$), showing that nonlinear relationships are especially relevant at this scale. For Immediate Regions with the network, EBM reaches $R^2 = 0.8196$, compared with $0.8112$ for the best linear model and $0.7788$ for the best Decision Tree.

\begin{table}[htbp]
\centering
\caption{Best $R^2$ by Configuration, Scale, and Algorithm}
\label{en:tab:melhores_geral}
\begin{tabular}{llccc}
\toprule
\textbf{Level} & \textbf{Variant} & \textbf{Best Linear} & \textbf{Best Tree} & \textbf{Best EBM} \\
\midrule
Municipality & Without Network & 0.7256 (Ridge) & 0.7277 & 0.7363 \\
Municipality & With Network & 0.7498 (LASSO) & 0.7647 & 0.7890 \\
Immediate Region & Without Network & 0.7893 (Ridge) & 0.7646 & 0.8050 \\
Immediate Region & With Network & 0.8112 (Ridge) & 0.7788 & \textbf{0.8196} \\
\bottomrule
\end{tabular}
\end{table}

\section{Conclusion}

This study investigated the predictive capacity of ICE and topological metrics from Brazil's transportation network for modeling IDHM at two levels of spatial aggregation: municipalities and Immediate Regions. The results address the three central research questions.

Regarding the best algorithm, EBM outperformed the linear models and Decision Trees in every configuration, reaching $R^2 = 0.8196$ for Immediate Regions in the With Network variant. This confirms that the relationships among economic complexity, the transportation network, and human development contain nonlinear components that linear regressors do not fully capture. The analysis of the ICE shape functions reinforces this conclusion: the relationship is not uniform, displaying volatile behavior at low ICE values, an almost linear central range, and diminishing returns above ICE = 1, a pattern that linear models approximate inadequately.

Regarding the effect of the transportation network, including topological metrics produced incremental gains across all models and scales. The improvement was most pronounced at the regional level, where the $R^2$ of the best linear model increased from 0.789 to 0.811. This result indicates that a locality's position in the transportation network is a relevant dimension of regional development, capturing access to markets and the logistical role of each territory.

Regarding territorial scale, the Immediate Region models produced consistently better results. At the municipality level, \texttt{reg\_Nordeste} dominated the models, revealing that the region's historical inequalities introduce a predictive component that ICE alone does not absorb at this scale. At the regional level, this dependence decreases and ICE becomes more relevant, confirming that Immediate Regions are the most coherent scale for analyzing the relationship between productive sophistication and human development.

Two directions stand out for future work. First, a deeper analysis of EBM shape functions, particularly for transportation-network metrics, may identify which centrality thresholds have the greatest impact on IDHM and thereby provide more concrete evidence for infrastructure policies. Second, investigating the influence of the Northeast variable may clarify the extent to which the lower development of municipalities in the region can be explained by productive sophistication rather than by other historical and structural determinants.

\section*{Appendix A: Detailed Grid-Search Results}

The following tables report the complete grid-search results for the Decision Trees and support the hyperparameter choices discussed in the main text.

\begin{table}[htbp]
\centering
\caption{Decision Tree Grid-Search Results---Municipalities}
\label{en:tab:grid_municipios}
\begin{tabular}{llccc}
\toprule
\textbf{Variant} & \textbf{max\_depth} & \textbf{min\_leaf} & \textbf{MAE} & \textbf{$R^2$} \\
\midrule
With Network & 7 & 10 & 0.0270 & 0.7647 \\
With Network & 7 & 2  & 0.0272 & 0.7613 \\
With Network & 7 & 5  & 0.0272 & 0.7609 \\
With Network & 5 & 10 & 0.0278 & 0.7538 \\
With Network & 5 & 2  & 0.0279 & 0.7537 \\
With Network & 5 & 5  & 0.0279 & 0.7533 \\
Without Network & 5 & 2  & 0.0294 & 0.7277 \\
Without Network & 5 & 5  & 0.0294 & 0.7285 \\
Without Network & 5 & 10 & 0.0294 & 0.7286 \\
Without Network & 7 & 10 & 0.0297 & 0.7219 \\
Without Network & 7 & 5  & 0.0298 & 0.7193 \\
Without Network & 7 & 2  & 0.0299 & 0.7151 \\
\bottomrule
\end{tabular}
\end{table}

\begin{table}[htbp]
\centering
\caption{Decision Tree Grid-Search Results---Immediate Regions}
\label{en:tab:grid_regioes_imediatas}
\begin{tabular}{llccc}
\toprule
\textbf{Variant} & \textbf{max\_depth} & \textbf{min\_leaf} & \textbf{MAE} & \textbf{$R^2$} \\
\midrule
With Network & 5 & 10 & 0.0247 & 0.7788 \\
With Network & 5 & 5  & 0.0250 & 0.7725 \\
With Network & 5 & 2  & 0.0253 & 0.7698 \\
With Network & 7 & 10 & 0.0253 & 0.7686 \\
With Network & 7 & 5  & 0.0262 & 0.7508 \\
With Network & 7 & 2  & 0.0272 & 0.7229 \\
Without Network & 7 & 10 & 0.0254 & 0.7646 \\
Without Network & 5 & 10 & 0.0255 & 0.7661 \\
Without Network & 5 & 5  & 0.0256 & 0.7614 \\
Without Network & 5 & 2  & 0.0261 & 0.7579 \\
Without Network & 7 & 5  & 0.0265 & 0.7430 \\
Without Network & 7 & 2  & 0.0284 & 0.6996 \\
\bottomrule
\end{tabular}
\end{table}

\clearpage
\putbib[referencias]
\end{bibunit}

\clearpage
\setcounter{section}{0}
\setcounter{subsection}{0}
\setcounter{figure}{0}
\setcounter{table}{0}
\setcounter{equation}{0}
\selectlanguage{brazil}
\renewcommand{\theHsection}{pt.\arabic{section}}
\renewcommand{\theHfigure}{pt.\arabic{figure}}
\renewcommand{\theHtable}{pt.\arabic{table}}
\begin{bibunit}[plain]
\begin{center}
{\LARGE\bfseries Complexidade econômica como determinante do desenvolvimento humano regional no Brasil:\\
Evidências por escala de agregação\par}
\vspace{1.5em}
{\large Eduardo Moura Zampirolli \quad Ruben Interian\par}
\vspace{0.6em}
Universidade Estadual de Campinas (Unicamp)\\
Instituto de Computação\\[0.6em]
2026
\end{center}
\vspace{1.5em}

\begin{abstract}
Nesta pesquisa, busca-se investigar a capacidade preditiva de modelos baseados na Complexidade Econômica e na Teoria de Redes Complexas aplicados ao Índice de Desenvolvimento Humano Municipal (IDHM) do Brasil. Para isso, o Índice de Complexidade Econômica (ICE) foi adaptado ao contexto brasileiro. O objetivo principal é avaliar em que medida diferentes métodos de análise do desenvolvimento de uma região, como a sua sofisticação produtiva local ou a inserção estrutural na malha de transportes nacional, determinam o desenvolvimento socioeconômico. A metodologia baseia-se na comparação de modelos de regressão lineares (Ridge, LASSO e ElasticNet) e não lineares (Árvores de Decisão e \textit{Explainable Boosting Machine}), submetidos à validação cruzada 5-fold com busca em grade (\textit{Grid Search}). As análises foram feitas em dois níveis de agregação espacial: municipal e de regiões imediatas. 
Os resultados revelam que os modelos agregados regionalmente apresentam menor ruído estatístico e maior estabilidade, alcançando coeficientes de determinação significativamente superiores, além de maior poder explicativo do ICE em relação às demais variáveis. Notavelmente, no nível de Regiões Imediatas, o ICE, isoladamente, desponta como o principal determinante estrutural do IDHM, sugerindo que a sofisticação produtiva regional explica, por si só, grande parte da variação do desenvolvimento humano nessa escala territorial. Ademais, a inclusão de métricas topológicas da rede de infraestrutura viária aumentou o desempenho de maneira geral. O modelo \textit{Explainable Boosting Machine} destacou-se como o algoritmo de melhor desempenho preditivo, alcançando $R^2 = 0{,}8196$ na escala de Regiões Imediatas com a inclusão das métricas de rede.

\end{abstract}

\section{Introdução}

A fim de auxiliar no planejamento estratégico e nas políticas públicas, é preciso compreender quais são os fatores cruciais para o desenvolvimento humano e econômico regional. Tradicionalmente, índices socioeconômicos isolados são utilizados para essa análise, porém fornecem uma visão limitada sobre as causas e estruturas que impulsionam o crescimento econômico. Com base nisso, a teoria da Complexidade Econômica propõe que o desenvolvimento está atrelado à sofisticação produtiva e às capacidades tecnológicas de um determinado local, que estão indicadas no Índice de Complexidade Econômica (ICE) \cite{pt:hidalgo2009}. 

Além disso, este trabalho adota os resultados de Morais (2025) \cite{pt:morais2025}, que modelou a infraestrutura rodoviária e hidroviária nacional como uma rede complexa ponderada para investigar a relação entre a posição geográfica dos municípios e seus índices de bem-estar. Isso permite testar uma hipótese integrada: o IDHM de uma localidade é simultaneamente determinado pela sua sofisticação produtiva interna (ICE) e pela sua facilidade de interação externa com outros mercados (métricas de rede viária). 

\section{Fundamentação Teórica}

Esta seção apresenta as bases teóricas para a modelagem e a construção das variáveis preditivas. A discussão é organizada em duas frentes: a Seção~\ref{sec:complexidade} trata dos conceitos de Complexidade Econômica e de suas métricas de sofisticação produtiva, enquanto a Seção~\ref{sec:rede_viaria_teoria} aborda a modelagem da infraestrutura de transportes pela ótica da Teoria de Redes Complexas.

\subsection{Complexidade Econômica}
\label{sec:complexidade}

Na formulação original de Hidalgo e Hausmann \cite{pt:hidalgo2009}, dados de exportações internacionais são usados para calcular o índice de complexidade econômica de uma nação. No entanto, aplicar essa métrica diretamente ao contexto municipal e regional do Brasil apresenta limitações metodológicas. A pauta de exportação brasileira é historicamente concentrada em \textit{commodities} e produtos primários de baixo valor agregado, o que não reflete a totalidade da capacidade produtiva do país. Como a economia nacional é fortemente orientada para o mercado interno e grande parte do avanço tecnológico e estrutural está no setor de serviços, o volume de exportações não capta a verdadeira complexidade dos municípios.

Por essa razão, este trabalho adota a adaptação ao contexto brasileiro validada por Monea \cite{pt:Monea2020}, que usa dados de emprego formal no lugar das exportações para mapear o conhecimento produtivo local. A base de dados primária é a Relação Anual de Informações Sociais (RAIS), que fornece o número de empregados formais por município e por atividade econômica, segundo a Classificação Nacional de Atividades Econômicas (CNAE). A partir desses dados, forma-se a matriz de especialização $X_{cp}$, cujos elementos representam o número de empregados na atividade $p$ no município $c$. 

Com base nessa matriz, calcula-se a Vantagem Comparativa Revelada (VCR), também denominada Quociente Locacional (QL), que mensura o grau de especialização relativa de cada município em cada atividade:
$$R_{cp} = \frac{X_{cp} \cdot X_{\text{Total}}}{X_c \cdot X_p}$$
onde $X_{\text{Total}} = \sum_{c,p} X_{cp}$ é o total de empregados na matriz, $X_c = \sum_p X_{cp}$ é o total de empregados no município $c$ e $X_p = \sum_c X_{cp}$ é o total de empregados na atividade $p$. A partir dessa razão, obtém-se a matriz binária de especialização $M_{cp}$, definida por:
$$M_{cp} = \begin{cases} 1, & \text{se } R_{cp} \geq 1 \\ 0, & \text{caso contrário} \end{cases}$$
Um município é considerado especializado em uma atividade quando sua participação relativa supera a média dos demais municípios, ou seja, quando $R_{cp} \geq 1$.

O Índice de Complexidade Econômica (ICE) é calculado pelo método do autovetor, seguindo a formulação de Hidalgo e Hausmann \cite{pt:hidalgo2009} e sua adaptação ao contexto brasileiro \cite{pt:Monea2020}. O método parte da construção de uma matriz de co-ocorrência entre municípios, $\tilde{M}_{cc'}$, que captura a similaridade das estruturas produtivas com base nas atividades que dois municípios compartilham:
$$\tilde{M}_{cc'} = \sum_p \frac{M_{cp} M_{c'p}}{k_{c,0} k_{p,0}}$$
onde $k_{c,0} = \sum_p M_{cp}$ é a \textit{diversidade} do município $c$ (número de atividades nas quais é especializado) e $k_{p,0} = \sum_c M_{cp}$ é a \textit{ubiquidade} da atividade $p$ (número de municípios especializados nela). Por construção, $\tilde{M}_{cc'}$ é uma matriz estocástica, de forma que seu autovetor dominante é um vetor constante e não informativo. O vetor de complexidade $K_c$ é o autovetor associado ao segundo maior autovalor de $\tilde{M}_{cc'}$, pois este captura a principal dimensão não trivial da estrutura produtiva.

O ICE é a versão padronizada (z-score) desse autovetor:
$$\text{ICE}_c = \frac{K_c - \langle K_c \rangle}{\sigma(K_c)}$$

A variável \textit{Diversity} corresponde diretamente a $k_{c,0}$, o número de atividades nas quais o município detém vantagem comparativa revelada.

\subsection{Infraestrutura Viária como Rede Complexa}
\label{sec:rede_viaria_teoria}

Para capturar a estrutura relacional do território nacional, o modelo utiliza métricas de centralidade. A partir da modelagem de uma rede em que os vértices são os municípios e as arestas são as vias de transporte ponderadas pelo tempo mínimo de deslocamento, extraem-se as seguintes variáveis topológicas, conforme Morais (2025) \cite{pt:morais2025}:
\begin{itemize}
    \item \textbf{Degree Centrality (Centralidade de Grau):} Mensura o número de conexões rodoviárias e hidroviárias diretas de cada município. Reflete o nível de conectividade local.
    \item \textbf{Betweenness Centrality (Centralidade de Intermediação):} Calcula a frequência com que um município aparece nos caminhos de menor tempo entre pares de outros municípios. Indica o poder do município como intermediário logístico.
    \item \textbf{Closeness Centrality (Centralidade de Proximidade):} Computa o inverso da distância média (em tempo de viagem) de um município em relação a todos os demais. Municípios com alta \textit{closeness} têm maior acessibilidade estrutural global.
    \item \textbf{Eigenvector Centrality (Centralidade de Autovetor):} Medida recursiva de importância na qual um município pontua alto se estiver conectado a outros polos centrais.
    \item \textbf{Average Neighbor Degree (Grau Médio dos Vizinhos):} Representa a média do número de conexões dos vizinhos diretos de um município.
    \item \textbf{Clustering Coefficient (Coeficiente de Agrupamento):} Avalia a conectividade entre os vizinhos de um município, indicando a existência de circuitos rodoviários locais coesos.
    \item \textbf{K-Core:} A decomposição em k-núcleos identifica o maior subgrafo em que todos os nós possuem pelo menos $k$ conexões entre si. O \textit{k-core} de um nó é o maior $k$ para o qual ele pertence a esse subgrafo, capturando sua inserção em regiões densamente conectadas da rede.
\end{itemize}

O trabalho de Morais evidenciou que essas variáveis possuem correlação estatisticamente significativa com o IDHM, principalmente \textit{Degree} e \textit{Closeness Centrality}, justificando seu uso em modelos preditivos mais complexos. 

\section{Objetivo}

A finalidade deste estudo é avaliar e comparar o desempenho de algoritmos de aprendizado de máquina explicáveis para modelar o IDHM brasileiro. Busca-se mensurar o ganho preditivo decorrente da inclusão de métricas de redes viárias aos modelos tradicionais compostos exclusivamente por dados socioeconômicos, validando a abordagem tanto no nível dos municípios quanto em estruturas macro (Regiões Imediatas).

Para isso, foram estabelecidos três objetivos específicos. Primeiro, identificar o algoritmo preditivo com a melhor capacidade de generalização e o menor erro estatístico. Segundo, analisar o desempenho do ICE e das métricas de rede viária em comparação com os demais preditores, construindo equivalências estatísticas do impacto de cada variável no IDHM. Por fim, determinar qual escala territorial, municipal ou por Regiões Imediatas, é mais adequada para capturar as dinâmicas da complexidade econômica. 

\section{Dados e Metodologia}

Esta seção detalha as bases de dados e as metodologias aplicadas, garantindo transparência analítica e reprodutibilidade dos experimentos. A estruturação abrange desde a coleta dos dados brutos até o treinamento dos modelos preditivos. Primeiro, são descritas as fontes primárias de dados. Na sequência, apresenta-se a construção do ICE e das métricas topológicas da rede viária. Depois, explica-se a estratégia de agregação espacial e o tratamento das heterogeneidades regionais. Por fim, são apresentados os modelos de aprendizado de máquina adotados e a estratégia de validação.

\subsection{Fontes de Dados}

Os dados da rede viária foram obtidos da base ``Ligações Rodoviárias e Hidroviárias do Brasil'', publicada pelo IBGE para o ano de 2016 \cite{pt:ibge2016}. Os indicadores socioeconômicos alvo (IDHM e seus componentes de Renda, Educação e Longevidade) foram coletados do Atlas do Desenvolvimento Humano no Brasil, referentes ao ano de 2010 \cite{pt:atlas2013}. Os dados de emprego formal para o cálculo do ICE foram obtidos da RAIS para o ano de 2020 \cite{pt:rais2020}.

\subsection{Construção dos Índices de Complexidade Econômica}

Conforme descrito na Seção~\ref{sec:complexidade}, o ICE foi calculado para todos os municípios brasileiros a partir da matriz binária de especialização $M_{cp}$, derivada da RAIS. A variável \textit{Diversity} ($k_{c,0}$), apesar de sua relevância teórica, apresenta alta correlação com o ICE. Ambas derivam da mesma matriz $M_{cp}$, e o ICE pode ser interpretado como uma medida de sofisticação ponderada pela diversidade. Por essa razão, e para garantir consistência entre os dois níveis de análise, essa variável foi excluída do conjunto de preditores.

\subsection{Métricas Topológicas da Rede Viária}

A rede de infraestrutura nacional foi modelada com os municípios como vértices e as rodovias e hidrovias como arestas ponderadas pelo tempo de deslocamento, utilizando a biblioteca \textit{NetworkX}. Para o nível de Regiões Imediatas, o grafo foi construído agregando as conexões municipais: uma aresta entre duas regiões é criada quando existe ao menos uma ligação direta entre municípios pertencentes a elas, com peso definido pela soma dos tempos de deslocamento de todas essas ligações, representando o volume total de conectividade entre os territórios.

Das métricas descritas na Seção~\ref{sec:rede_viaria_teoria}, foram selecionadas como \textit{features} preditivas finais: \textit{Degree Centrality}, \textit{Betweenness Centrality}, \textit{Closeness Centrality}, \textit{Eigenvector Centrality}, \textit{K-Core}, \textit{Average Neighbor Degree} e \textit{Clustering Coefficient}.

\subsection{Integração e Níveis de Agregação Espacial}

Os dados foram organizados em dois níveis espaciais. No nível de Municípios, os preditores são o ICE e as métricas de rede. No nível de Regiões Imediatas, as variáveis municipais foram consolidadas por médias ponderadas pela população de 2010.

Para capturar as diferenças estruturais históricas entre as macrorregiões brasileiras, como desigualdades na industrialização e nas políticas de desenvolvimento, foram incluídas variáveis binárias (\textit{dummy variables}) indicadoras das cinco macrorregiões. Adotando o Centro-Oeste como categoria de referência, foram incluídas \texttt{reg\_Norte}, \texttt{reg\_Nordeste}, \texttt{reg\_Sudeste} e \texttt{reg\_Sul} em todas as variantes e níveis de agregação.

\subsection{Modelos Preditivos e Estratégia de Validação}

Os modelos testados foram: Regressões Lineares Regularizadas (Ridge, LASSO e ElasticNet), Árvores de Decisão e \textit{Explainable Boosting Machine} (EBM). Valores ausentes nas \textit{features} foram imputados pela mediana de cada variável, calculada sobre o conjunto de treinamento em cada \textit{fold}. O ajuste fino foi realizado via busca em grade (\textit{Grid Search}) com validação cruzada de 5 \textit{folds}, usando a minimização do Erro Absoluto Médio (MAE) como critério principal.

\subsection{Explainable Boosting Machine (EBM)}
\label{sec:ebm_metodo}

O \textit{Explainable Boosting Machine} (EBM) é um modelo aditivo generalizado (GAM) baseado em \textit{gradient boosting}, proposto por Lou et al. \cite{pt:lou2012} e implementado na biblioteca \textit{InterpretML} \cite{pt:nori2019interpretml}. Sua formulação é:
$$\hat{y} = \beta_0 + \sum_i f_i(x_i) + \sum_{i < j} f_{ij}(x_i, x_j)$$
onde cada $f_i$ é uma função de forma aprendida para a variável $x_i$ e os termos $f_{ij}$ capturam interações par a par. Ao contrário das Árvores de Decisão, o EBM mantém interpretabilidade total: a contribuição individual de cada variável pode ser visualizada separadamente, o que é especialmente útil para análises de políticas públicas. Os hiperparâmetros fixados foram \textit{inner\_bags} = 5 e \textit{outer\_bags} = 50; o parâmetro otimizado via \textit{Grid Search} foi o número de interações pareadas, avaliado em $\{0, 5, 10, 15\}$.

\section{Resultados e Discussão}

Esta seção apresenta os resultados obtidos. Primeiro, são analisados os três modelos testados (lineares, Árvores de Decisão e EBM). Em seguida, avalia-se o impacto da inclusão da rede viária e o efeito da escala de agregação espacial. Por fim, discute-se a influência da variável regional do Nordeste e a importância relativa das variáveis, com uma síntese comparativa ao final.

\subsection{Modelos Lineares Regularizados}

Os regressores LASSO, Ridge e ElasticNet apresentaram desempenhos muito próximos entre si nos dois níveis de agregação, conforme as Tabelas~\ref{tab:cv_municipios} e~\ref{tab:cv_regioes}. Essa convergência indica que o conjunto de preditores não apresenta multicolinearidade severa. No nível municipal, as diferenças entre os três modelos ficaram na ordem de $10^{-4}$ tanto no MAE quanto no $R^2$. No nível regional, o Ridge apresentou leve vantagem no $R^2$ (diferença da ordem de $10^{-3}$). O comportamento similar entre os três regressores sugere que as relações capturadas são predominantemente lineares, o que motiva o uso de modelos não-lineares nas seções seguintes.

\subsection{Árvores de Decisão}

Conforme a Tabela~\ref{tab:resultados_arvore}, a Árvore de Decisão para Regiões Imediatas (Com Rede) atingiu seu melhor resultado com \textit{max\_depth} = 5 e \textit{min\_leaf} = 10 (MAE = 0,0247). No nível municipal, a profundidade ideal foi 7, refletindo a maior heterogeneidade das amostras nessa escala.

Nos gráficos de redução de impureza (Figura~\ref{fig:red_impur_arvores}), o comportamento é distinto entre os dois níveis. No nível municipal, \texttt{reg\_Nordeste} supera o ICE como preditor principal, com importância acima de 0,50 em ambas as variantes Com e Sem Rede. No nível de Regiões Imediatas, o ICE domina, com valores acima de 0,90, e \texttt{reg\_Nordeste} perde relevância. As métricas de rede viária tiveram importâncias individuais baixas, mas sua inclusão gerou ganhos consistentes no $R^2$, especialmente no nível municipal.

\subsection{Explainable Boosting Machine (EBM)}


O EBM obteve o melhor desempenho entre todos os modelos testados. Na escala de Regiões Imediatas com a variante Com Rede, o modelo alcançou MAE = 0,0226 e $R^2 = 0,8196$ --- os melhores valores da pesquisa, conforme a Tabela~\ref{tab:ebm_resultados}. O ganho sobre os modelos lineares se deve à capacidade do EBM de capturar relações não-lineares e interações entre variáveis de forma automatizada. O ajuste com 5 a 15 interações provou-se o mais equilibrado entre poder preditivo e generalização.

\begin{figure}[!htbp]
    \centering
    \begin{subfigure}[b]{0.48\textwidth}
        \includegraphics[width=\textwidth]{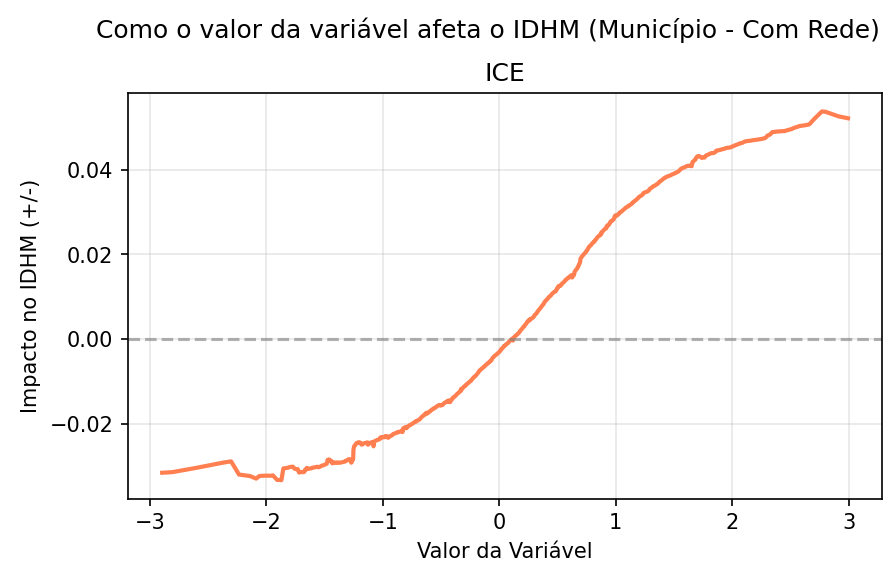}
        \caption{Municípios com rede viária}
        \label{fig:shape_mun_rede}
    \end{subfigure}
    \hfill
    \begin{subfigure}[b]{0.48\textwidth}
        \includegraphics[width=\textwidth]{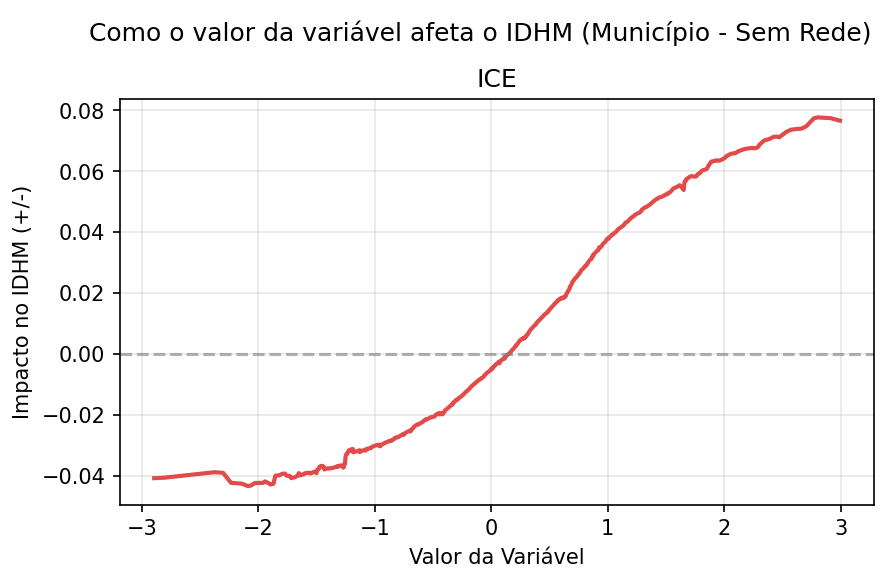}
        \caption{Municípios sem rede viária}
        \label{fig:shape_mun_sem}
    \end{subfigure}

    \vspace{1em}

    \begin{subfigure}[b]{0.48\textwidth}
        \includegraphics[width=\textwidth]{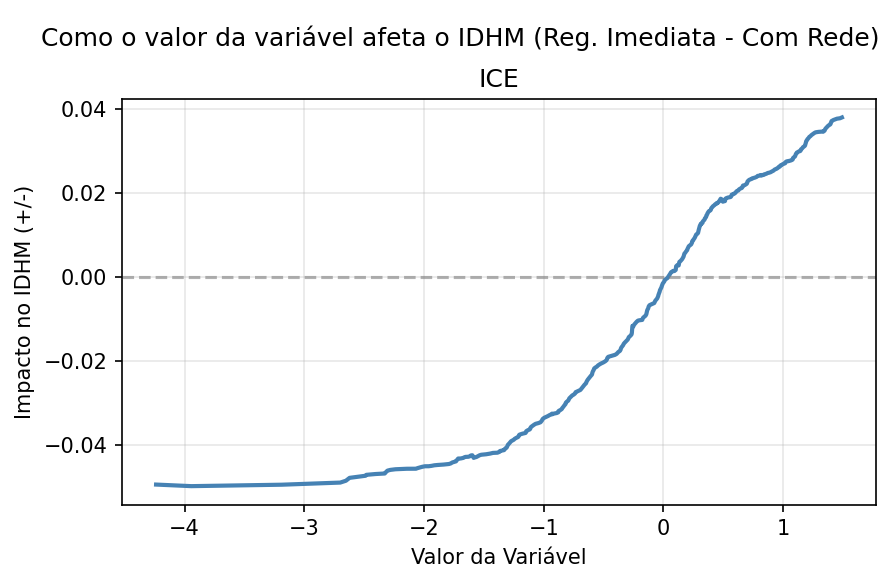}
        \caption{Regiões Imediatas com rede viária}
        \label{fig:shape_imm_rede}
    \end{subfigure}
    \hfill
    \begin{subfigure}[b]{0.48\textwidth}
        \includegraphics[width=\textwidth]{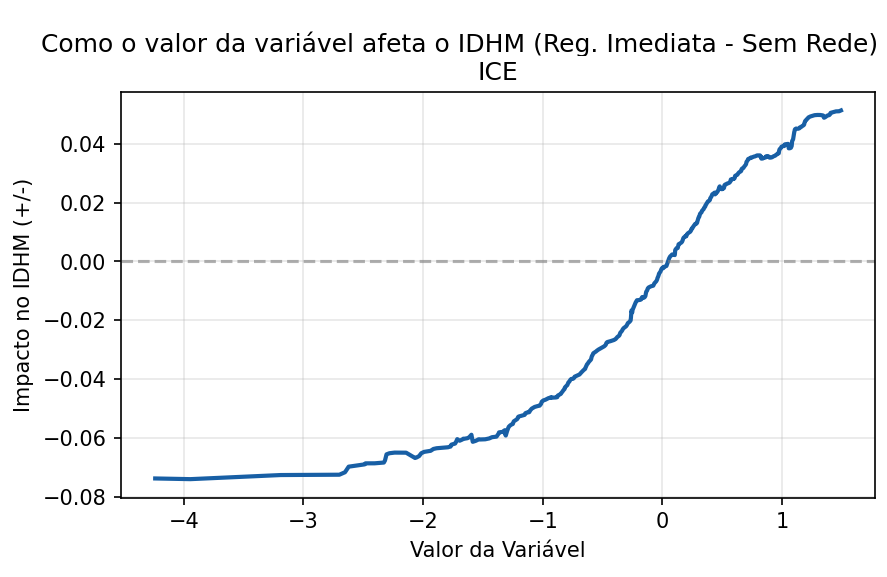}
        \caption{Regiões Imediatas sem rede viária}
        \label{fig:shape_imm_sem}
    \end{subfigure}

    \caption{Funções de forma (\textit{shape functions}) do ICE no EBM,
    por nível de agregação e variante. O eixo vertical representa o impacto
    aditivo no IDHM previsto; o eixo horizontal representa o valor do ICE.}
    \label{fig:shape_ice}
\end{figure}

As funções de forma do ICE (Figura~\ref{fig:shape_ice}) confirmam essa não-linearidade: a relação com o IDHM é volátil e de baixa inclinação para ICE muito negativo, cresce de forma quase linear na faixa central ($-1$ a $1$) e apresenta retornos decrescentes acima de $1$. No nível de Regiões Imediatas, o mesmo padrão aparece com menor variabilidade nas extremidades, reflexo do efeito suavizador da agregação espacial. Em todas as configurações, a amplitude da curva é menor na variante Com Rede, indicando que as métricas de rede absorvem parte da variação do IDHM antes atribuída exclusivamente ao ICE.

No que se refere à importância relativa das variáveis (Figura~\ref{fig:red_impur_ebm}), o padrão das Árvores de Decisão se repete: no nível municipal, \texttt{reg\_Nordeste} mantém-se como preditor de maior peso, enquanto o ICE lidera no nível regional. A diferença entre as duas variáveis, porém, é menor no EBM do que nas Árvores, indicando que o modelo aditivo distribui a explicação de forma mais equilibrada, aproveitando melhor as interações entre ICE, métricas de rede e indicadores regionais.

\begin{table}[htbp]
\centering
\caption{Comparação de Desempenho do Modelo EBM (Melhores Hiperparâmetros)}
\label{tab:ebm_resultados}
\begin{tabular}{llccc}
\toprule
\textbf{Nível} & \textbf{Variante} & \textbf{Interações Otimizadas} & \textbf{MAE} & \textbf{R²} \\
\midrule
Município & Sem Rede & 10 & 0.0289 & 0.7363 \\
Município & Com Rede & 15 & 0.0258 & 0.7890 \\
Reg. Imediata & Sem Rede & 5 & 0.0238 & 0.8050 \\
Reg. Imediata & Com Rede & 15 & 0.0226 & 0.8196 \\
\bottomrule
\end{tabular}
\end{table}

\subsection{Inclusão das Métricas de Rede na Análise}

Conforme observado nas seções anteriores, a inclusão das métricas de rede gera ganho sistemático de desempenho em ambos os níveis e em todas as famílias de modelos. No nível municipal, a variante Com Rede elevou o $R^2$ dos regressores lineares de aproximadamente 0,726 para 0,750. No nível de Regiões Imediatas, o efeito foi ainda mais expressivo: o melhor modelo linear passou de $R^2 = 0,789$ para $R^2 = 0,811$, conforme a Tabela~\ref{tab:melhores_geral}. Esse padrão crescente com a escala de agregação indica que as métricas de centralidade viária capturam dimensões do desenvolvimento especialmente relevantes no contexto regional, onde o acesso a mercados e a posição logística afetam o conjunto dos municípios de forma coletiva.

\subsection{Efeito da Agregação Espacial e Interdependência Produtiva}
Ao comparar as Tabelas~\ref{tab:cv_municipios} e~\ref{tab:cv_regioes} é possível observar uma diferença clara no desempenho entre a escala municipal e regional. Nos modelos lineares, o $R^2$ sobe de aproximadamente 0,73 para 0,79 na variante Sem Rede, e de 0,75 para 0,81 na variante Com Rede --- um ganho de 6 a 7 pontos percentuais simplesmente pela mudança de escala territorial. Para referência, a inclusão das métricas de rede viária dentro de cada escala gera um ganho de apenas 2 a 3 pontos percentuais. O EBM segue o mesmo padrão, passando de $R^2 = 0,789$ (municipal, Com Rede) para $R^2 = 0,820$ (regional, Com Rede). Isso significa que a escolha da escala de agregação impacta o desempenho preditivo mais do que qualquer outro fator metodológico testado nesta pesquisa.

Esse resultado não é apenas quantitativo: a mudança de escala também altera qualitativamente o papel do ICE nos modelos. Os diagramas de importância das Árvores de Decisão e do EBM (Figuras~\ref{fig:red_impur_arvores} e~\ref{fig:red_impur_ebm}) mostram que, no nível municipal, o ICE divide espaço com \texttt{reg\_Nordeste} como preditor principal, enquanto no nível de Regiões Imediatas o ICE responde sozinho por mais de 90\% da redução de impureza. Em outras palavras, a complexidade econômica passa a ser suficiente para explicar as variações no IDHM quando os dados estão agregados regionalmente, mas não quando estão no recorte local. Isso aponta para uma limitação estrutural da modelagem municipal: nessa escala, o ICE de um município isolado carrega informação insuficiente sobre o seu verdadeiro ambiente produtivo.

A razão disso está no próprio funcionamento das economias regionais. O IDHM de um município não depende apenas de sua estrutura produtiva interna, mas é fortemente moldado pela complexidade dos territórios ao seu redor, seja pelo acesso a mercados de trabalho mais sofisticados, pelo fluxo de capital e de serviços especializados, ou pela difusão de conhecimento técnico entre firmas próximas. Ao tentar modelar o desenvolvimento no recorte municipal, o algoritmo opera sem acesso a essas externalidades, o que se traduz em maior erro residual e menor capacidade explicativa do ICE. Ao agregar os municípios em Regiões Imediatas, essas interdependências passam a estar implicitamente representadas no próprio valor médio do ICE regional, que reflete a sofisticação do ecossistema produtivo como um todo e não apenas de cada unidade isolada.

\begin{figure}[!htbp]
    \centering
    \includegraphics[width=0.98\linewidth]{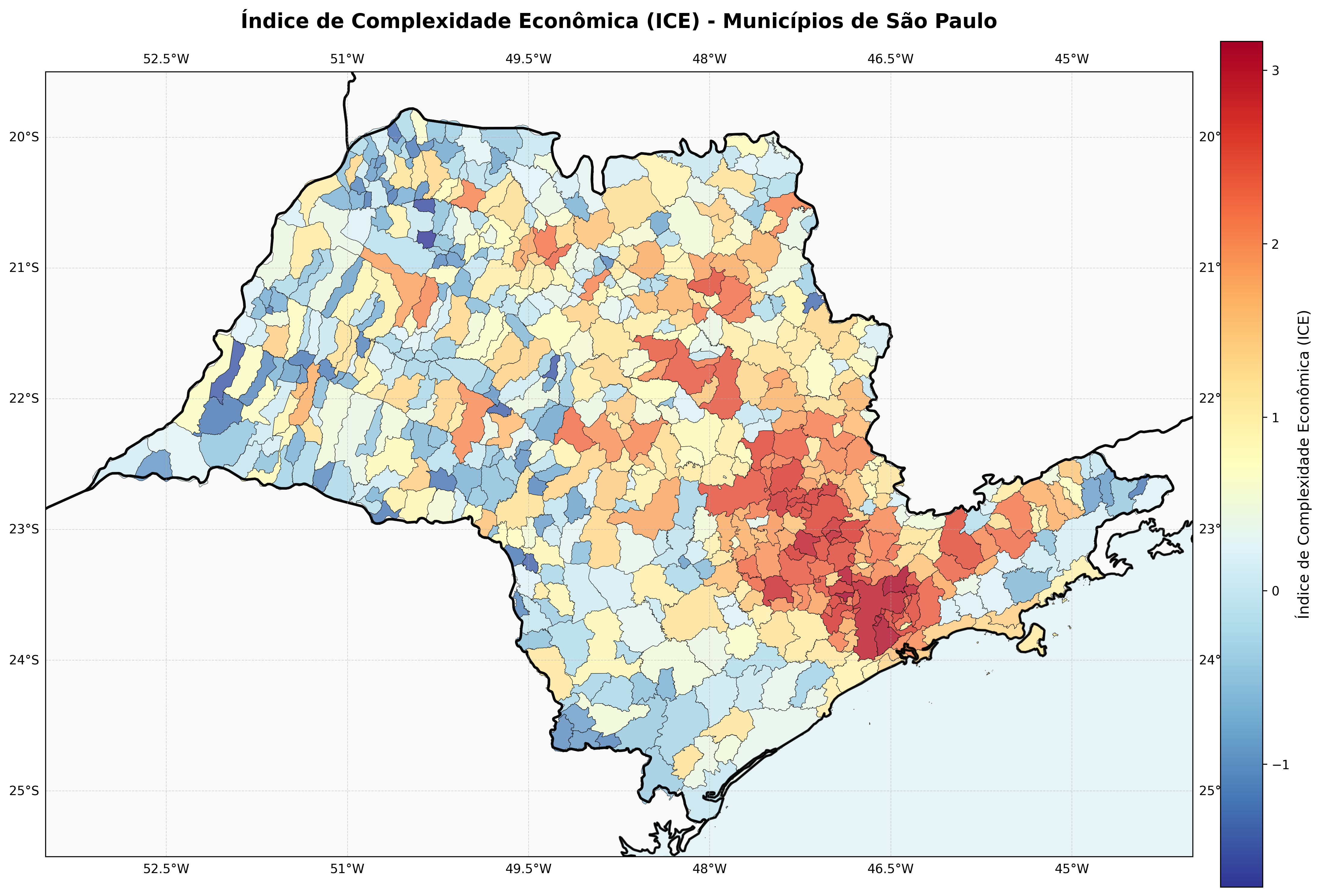}
    \caption{Mapa da complexidade econômica dos municípios do estado de São Paulo.}
    \label{fig:mapa_sp}
\end{figure}

Essa dinâmica é demonstrada pelo caso da Região Metropolitana de São Paulo, detalhado na Tabela~\ref{tab:ice_idhm_municipios} e visualizado na Figura~\ref{fig:mapa_sp}. A capital paulista apresenta IDHM = 0,805, porém seu ICE (2,949) é superado por municípios industriais do entorno: Guarulhos (ICE = 3,127, IDHM = 0,763), São Bernardo do Campo (ICE = 3,076, IDHM = 0,805) e Diadema (ICE = 3,024, IDHM = 0,757). Esses municípios concentram a produção de maior sofisticação tecnológica da região, mas apresentam IDHM igual ou inferior ao da capital, que se especializa em serviços, finanças e gestão. No nível municipal, esse padrão cria uma desconexão entre ICE e IDHM que o modelo não consegue explicar localmente: o município com maior ICE não é necessariamente o de maior IDHM. No entanto, ao agregar essas cidades em uma única Região Imediata, o ICE médio regional passa a refletir um ecossistema produtivo integrado, e a relação com o IDHM médio ponderado da região torna-se muito mais estável e interpretável. É justamente essa coerência estatística --- ausente no recorte municipal, presente no regional --- que explica o salto de desempenho observado nos modelos.

\begin{table}[htbp]
\centering
\caption{Índice de Complexidade Econômica (ICE) e IDHM por Município}
\label{tab:ice_idhm_municipios}
\begin{tabular}{llcc}
\toprule
\textbf{Município} & \textbf{ID Município} & \textbf{ICE} & \textbf{IDHM} \\
\midrule
São Paulo & 3550308 & 2.949147 & 0.805 \\
\midrule
Guarulhos & 3518800 & 3.127885 & 0.763 \\
São Bernardo do Campo & 3548708 & 3.076483 & 0.805 \\
Diadema & 3513801 & 3.024338 & 0.757 \\
São Caetano do Sul & 3548807 & 2.840713 & 0.862 \\
Cotia & 3513009 & 2.721974 & 0.780 \\
Mauá & 3529401 & 2.705295 & 0.766 \\
Santo André & 3547809 & 2.643223 & 0.815 \\
Taboão da Serra & 3552809 & 2.561624 & 0.769 \\
Poá & 3539806 & 2.507171 & 0.776 \\
Osasco & 3534401 & 2.485954 & 0.776 \\
Ferraz de Vasconcelos & 3515707 & 2.329641 & 0.738 \\
Juquitiba & 3525300 & 2.052169 & 0.709 \\
Mairiporã & 3528502 & 2.030977 & 0.788 \\
Itapecerica da Serra & 3522208 & 1.751917 & 0.742 \\
Embu-Guaçu & 3515103 & 1.566397 & 0.749 \\
Itaquaquecetuba & 3522406 & 0.837479 & 0.714 \\
\bottomrule
\end{tabular}
\end{table}

\begin{table}[htbp]
\centering
\caption{Comparação de Desempenho dos Modelos - Nível Município (CV 5-fold)}
\label{tab:cv_municipios}
\begin{tabular}{llcc}
\toprule
\textbf{Variante} & \textbf{Modelo} & \textbf{MAE} & \textbf{$R^2$} \\
\midrule
\multirow{3}{*}{Com Rede}  
 & LASSO      & 0.0281 & 0.7498 \\
 & ElasticNet & 0.0281 & 0.7498 \\
 & Ridge      & 0.0282 & 0.7490 \\
\midrule
\multirow{3}{*}{Sem Rede}  
 & Ridge      & 0.0294 & 0.7256 \\
 & LASSO      & 0.0294 & 0.7255 \\
 & ElasticNet & 0.0295 & 0.7254 \\
\bottomrule
\end{tabular}
\end{table}

\begin{table}[htbp]
\centering
\caption{Comparação de Desempenho dos Modelos - Nível Região Imediata (CV 5-fold)}
\label{tab:cv_regioes}
\begin{tabular}{llcc}
\toprule
\textbf{Variante} & \textbf{Modelo} & \textbf{MAE} & \textbf{$R^2$} \\
\midrule
\multirow{3}{*}{Com Rede}  
 & Ridge      & 0.0237 & 0.8112 \\
 & LASSO      & 0.0237 & 0.8087 \\
 & ElasticNet & 0.0237 & 0.8088 \\
\midrule
\multirow{3}{*}{Sem Rede}  
 & Ridge      & 0.0250 & 0.7893 \\
 & ElasticNet & 0.0251 & 0.7874 \\
 & LASSO      & 0.0251 & 0.7873 \\
\bottomrule
\end{tabular}
\end{table}

\begin{table}[htbp]
\centering
\caption{Resultados da Otimização da Árvore de Decisão por Nível e Variante}
\label{tab:resultados_arvore}
\begin{tabular}{lccccc}
\toprule
\textbf{Nível} & \textbf{Variante} & \textbf{max\_depth} & \textbf{min\_leaf} & \textbf{MAE} & \textbf{R²} \\
\midrule
Município     & Sem Rede & 5 & 2  & 0.0294 & 0.7277 \\
Município     & Com Rede & 7 & 10 & 0.0270 & 0.7647 \\
\midrule
Reg. Imediata & Sem Rede & 7 & 10 & 0.0254 & 0.7646 \\
Reg. Imediata & Com Rede & 5 & 10 & 0.0247 & 0.7788 \\
\bottomrule
\end{tabular}
\end{table}

\subsection{Influência Regional}

Um dos achados mais relevantes é o peso da variável \texttt{reg\_Nordeste} nos modelos, especialmente no nível municipal. Nas Árvores de Decisão (Figura~\ref{fig:red_impur_arvores}), essa variável supera o ICE como principal preditor do IDHM municipal, com importância para a redução de impureza acima de 0,50 em ambas as variantes. No EBM, essa predominância se mantém com menor magnitude.

Esse resultado reflete características estruturais da realidade brasileira: municípios nordestinos apresentam IDHM sistematicamente abaixo da média nacional por razões que vão além da sofisticação produtiva local. A região tem menor industrialização histórica, menor cobertura de infraestrutura, maior dependência de transferências governamentais e predominância de atividades agropecuárias extensivas. Em outras palavras, pertencer ao Nordeste agrega um fator preditivo sobre o IDHM que o ICE sozinho não captura no nível municipal.

Isso revela uma limitação do modelo local: no recorte dos municípios, o algoritmo tende a separar primeiro os ``municípios nordestinos'' dos demais, antes de usar a sofisticação produtiva como refinamento. Os primeiros nós de divisão das Árvores de Decisão funcionam essencialmente como um classificador regional, só depois incorporando o ICE e as métricas de rede.

No nível de Regiões Imediatas, esse efeito diminui bastante: o ICE passa a ser o preditor dominante e \texttt{reg\_Nordeste} perde relevância relativa. A agregação espacial dilui parte da heterogeneidade interna da região, e as diferenças estruturais do Nordeste passam a ser capturadas pela própria variação do ICE entre as regiões imediatas, que é consistentemente menor nessa macrorregião. Isso reforça que a escala de Regiões Imediatas é mais adequada para analisar a relação entre complexidade produtiva e desenvolvimento humano de forma integrada.

\subsection{Importância das Variáveis e Interpretabilidade}

\begin{figure}[!htbp]
    \centering
    \begin{subfigure}[b]{0.48\textwidth}
        \includegraphics[width=\textwidth]{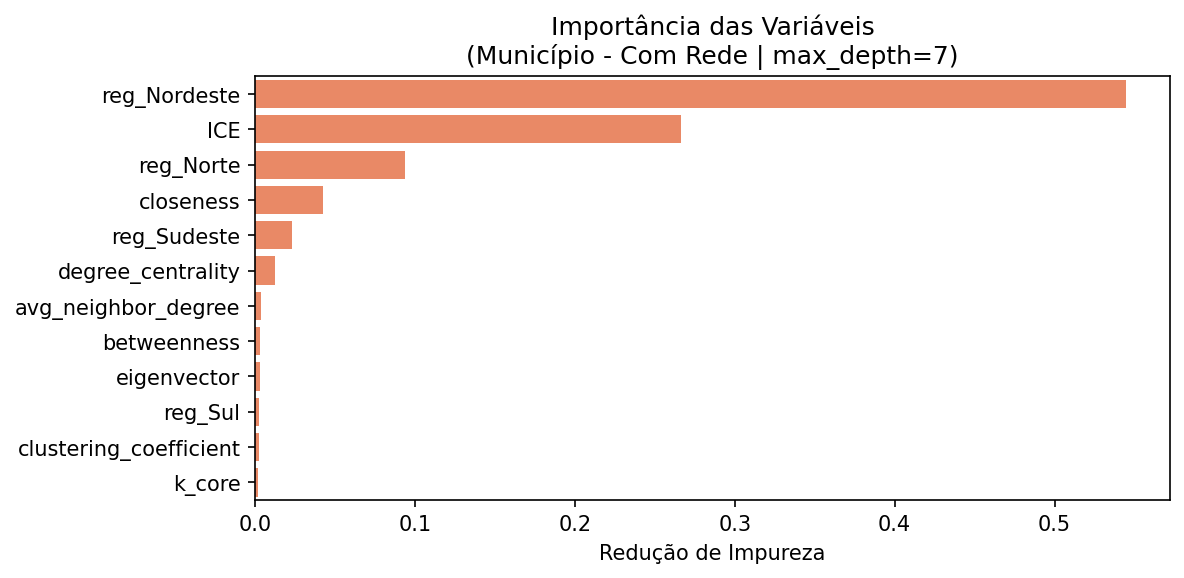}
        \caption{Municípios com rede viária}
        \label{fig:imp_mun_tree_rede}
    \end{subfigure}
    \hfill
    \begin{subfigure}[b]{0.48\textwidth}
        \includegraphics[width=\textwidth]{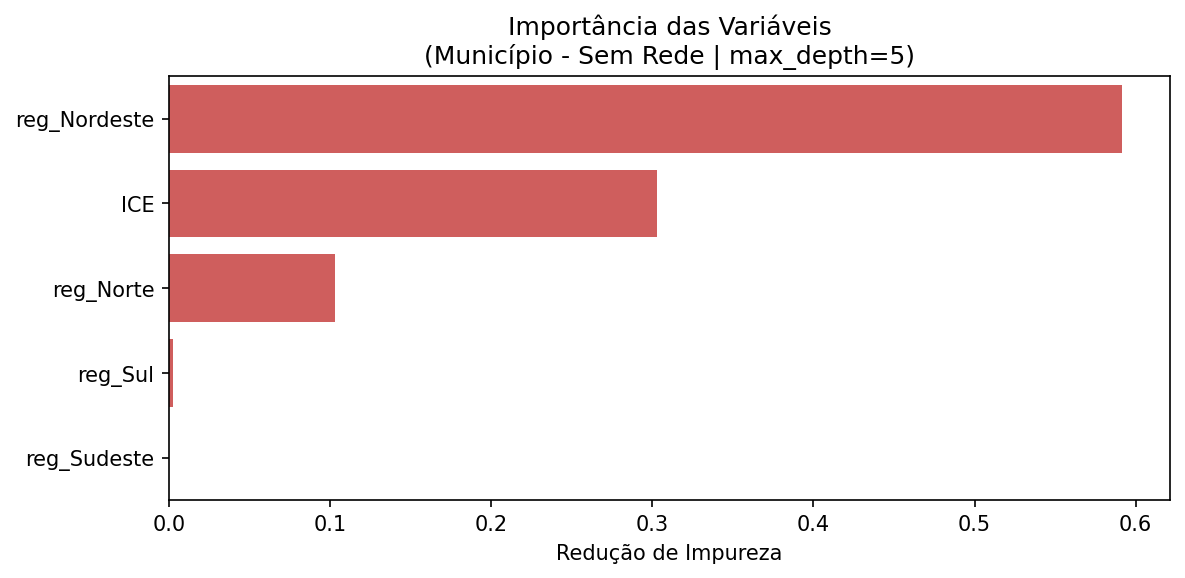}
        \caption{Municípios sem rede viária}
        \label{fig:imp_mun_tree_sem}
    \end{subfigure}
    
    \vspace{1em}
    
    \begin{subfigure}[b]{0.48\textwidth}
        \includegraphics[width=\textwidth]{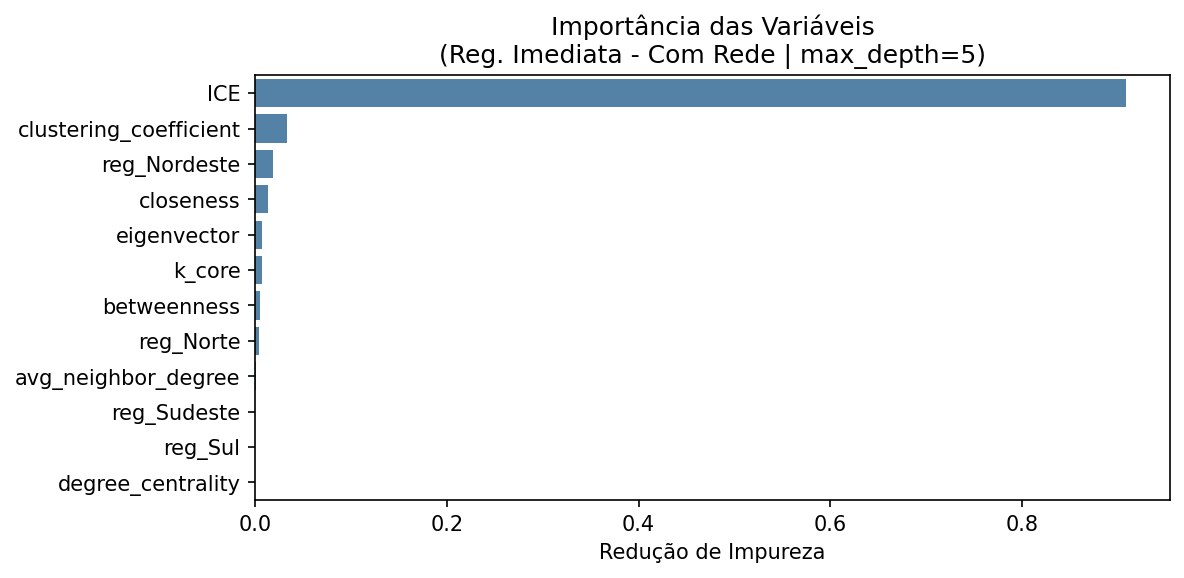}
        \caption{Regiões Imediatas com rede viária}
        \label{fig:imp_imm_tree_rede}
    \end{subfigure}
    \hfill
    \begin{subfigure}[b]{0.48\textwidth}
        \includegraphics[width=\textwidth]{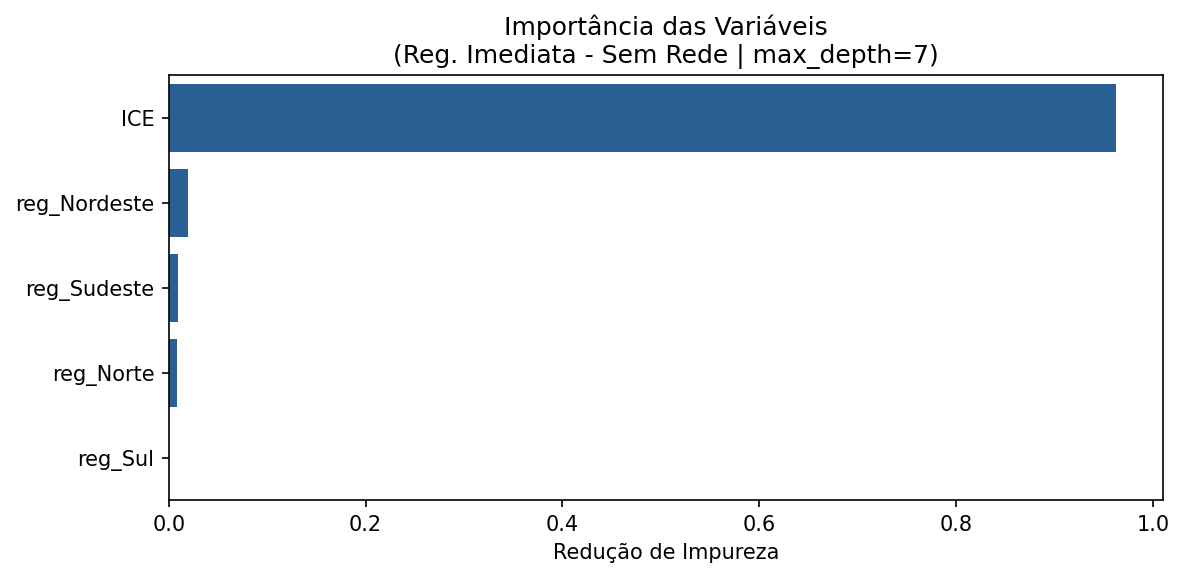}
        \caption{Regiões Imediatas sem rede viária}
        \label{fig:imp_imm_tree_sem}
    \end{subfigure}
    
    \caption{Redução de impureza das variáveis preditivas nas Árvores de Decisão, por nível de agregação e variante.}
    \label{fig:red_impur_arvores}
\end{figure}

\begin{figure}[!htbp]
    \centering
    \begin{subfigure}[b]{0.48\textwidth}
        \includegraphics[width=\textwidth]{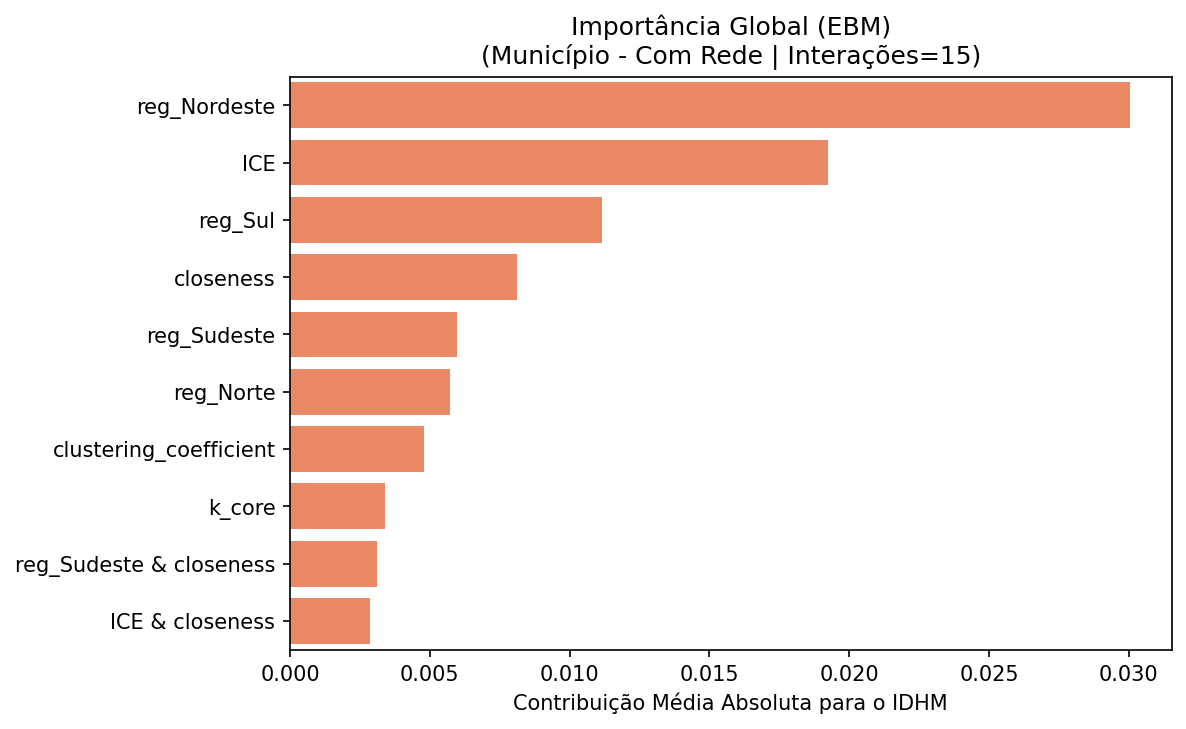}
        \caption{Municípios com rede viária}
        \label{fig:imp_mun_ebm_rede}
    \end{subfigure}
    \hfill
    \begin{subfigure}[b]{0.48\textwidth}
        \includegraphics[width=\textwidth]{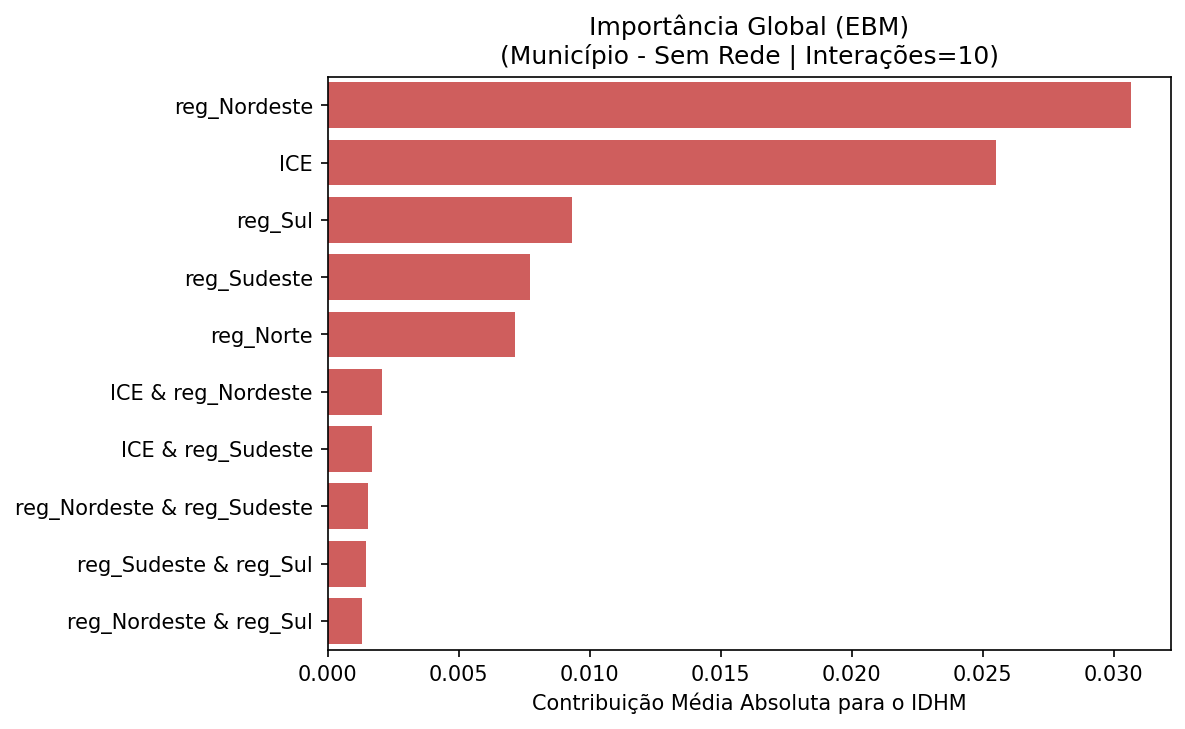}
        \caption{Municípios sem rede viária}
        \label{fig:imp_mun_ebm_sem}
    \end{subfigure}
    
    \vspace{1em}
    
    \begin{subfigure}[b]{0.48\textwidth}
        \includegraphics[width=\textwidth]{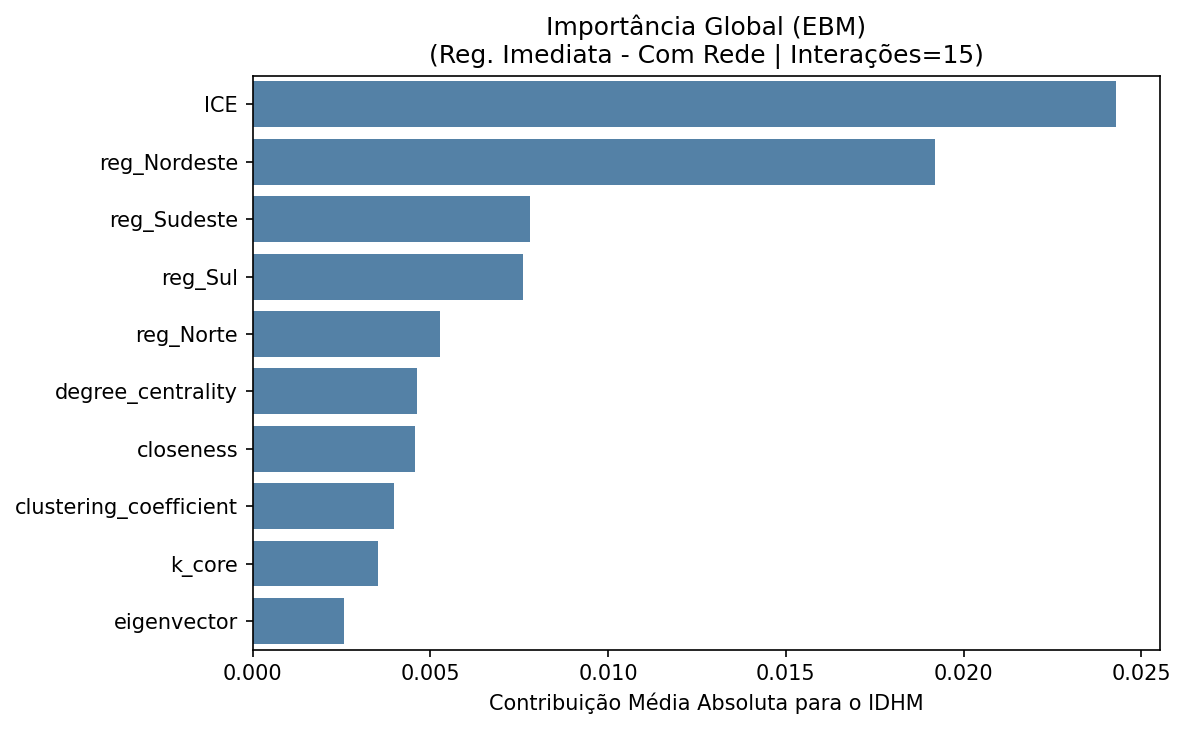}
        \caption{Regiões Imediatas com rede viária}
        \label{fig:imp_imm_ebm_rede}
    \end{subfigure}
    \hfill
    \begin{subfigure}[b]{0.48\textwidth}
        \includegraphics[width=\textwidth]{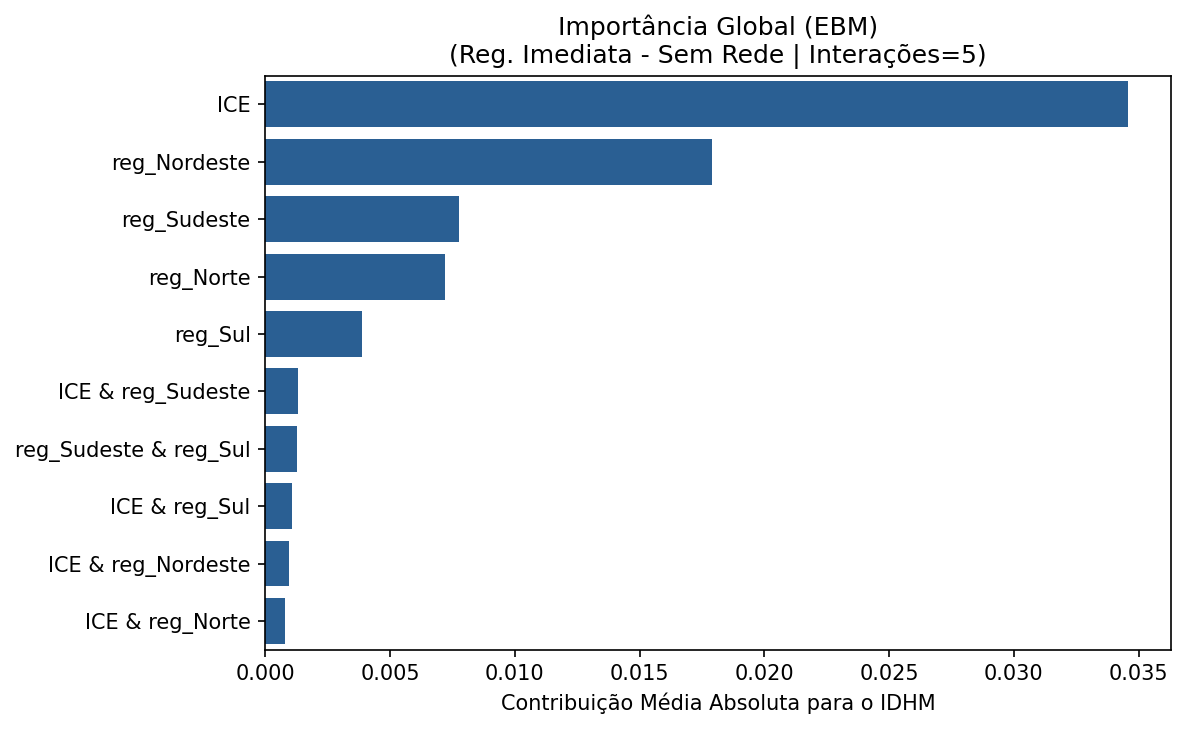}
        \caption{Regiões Imediatas sem rede viária}
        \label{fig:imp_imm_ebm_sem}
    \end{subfigure}
    
    \caption{Importância das variáveis preditivas no EBM, por nível de agregação e variante.}
    \label{fig:red_impur_ebm}
\end{figure}

Os diagramas de redução de impureza das Árvores de Decisão (Figura~\ref{fig:red_impur_arvores}) e de importância do EBM (Figura~\ref{fig:red_impur_ebm}) mostram um padrão consistente: no nível municipal, \texttt{reg\_Nordeste} é o preditor mais relevante, seguida pelo ICE; no nível de Regiões Imediatas, o ICE assume a liderança com ampla margem. Entre as métricas de rede, \textit{Closeness Centrality} e \textit{Degree Centrality} foram as mais consistentes nos dois níveis e modelos, ainda que com importâncias individuais menores que as variáveis socioeconômicas.

\subsection{Síntese Comparativa dos Modelos}

A Tabela~\ref{tab:melhores_geral} resume os melhores resultados de cada família de modelos por configuração. Dois padrões se destacam. Primeiro, a modelagem por Regiões Imediatas supera consistentemente a modelagem municipal em todas as combinações --- o ganho de $R^2$ ao mudar de escala é maior do que qualquer outro fator analisado. Segundo, o EBM foi o melhor modelo em todas as configurações, com margem mais expressiva no nível municipal Com Rede ($R^2 = 0,789$) em relação ao Linear ($R^2 = 0,750$) e à Árvore ($R^2 = 0,765$), mostrando que as relações não-lineares são especialmente relevantes nessa escala. No nível regional Com Rede, o EBM alcançou $R^2 = 0,8196$, ante $0,8112$ do melhor linear e $0,7788$ da melhor Árvore.

\begin{table}[htbp]
\centering
\caption{Síntese do Melhor Modelo ($R^2$) por Configuração, Escala e Algoritmo}
\label{tab:melhores_geral}
\begin{tabular}{llccc}
\toprule
\textbf{Nível} & \textbf{Variante} & \textbf{Melhor Linear} & \textbf{Melhor Árvore} & \textbf{Melhor EBM} \\
\midrule
Município     & Sem Rede & 0.7256 (Ridge) & 0.7277 & 0.7363 \\
Município     & Com Rede & 0.7498 (LASSO) & 0.7647 & 0.7890 \\
Reg. Imediata & Sem Rede & 0.7893 (Ridge) & 0.7646 & 0.8050 \\
Reg. Imediata & Com Rede & 0.8112 (Ridge) & 0.7788 & \textbf{0.8196} \\
\bottomrule
\end{tabular}
\end{table}

\section{Conclusão}

Este trabalho investigou a capacidade preditiva do ICE e de métricas topológicas da rede viária nacional para modelar o IDHM do Brasil em dois níveis de agregação espacial: municípios e Regiões Imediatas. Os resultados permitem responder às três questões centrais da pesquisa.

Quanto ao melhor algoritmo, o EBM superou os modelos lineares e as Árvores de Decisão em todas as configurações, alcançando $R^2 = 0,8196$ (Regiões Imediatas, Com Rede). Isso confirma que as relações entre complexidade econômica, rede viária e desenvolvimento humano têm componentes não-lineares que os regressores lineares não capturam totalmente. A análise das funções de forma do ICE reforça essa conclusão: a relação não é uniforme, com comportamento volátil em valores baixos de ICE, zona quase linear na faixa central e retornos decrescentes acima de ICE = 1, padrão que os modelos lineares aproximam de forma insatisfatória.

Quanto ao efeito da rede viária, a inclusão das métricas topológicas gerou ganho incremental em todos os modelos e escalas, com melhora mais expressiva no nível regional, o $R^2$ do melhor modelo linear passou de 0,789 para 0,811. Esse resultado indica que a posição de uma localidade na malha de transportes é uma dimensão relevante do desenvolvimento regional, capturando o acesso a mercados e o papel logístico de cada território.

Quanto à escala territorial, os modelos no nível de Regiões Imediatas produziram resultados consistentemente superiores. No nível municipal, a variável \texttt{reg\_Nordeste} dominou os modelos, revelando que desigualdades históricas da região introduzem um componente preditivo que o ICE não absorve sozinho nessa escala. No nível regional, essa dependência diminui e o ICE assume maior relevância, confirmando que as Regiões Imediatas são a escala mais coerente para analisar a relação entre sofisticação produtiva e desenvolvimento humano.

Como trabalhos futuros, duas direções se destacam. Em primeiro lugar, o aprofundamento da análise das funções de forma do EBM, em particular nas métricas de rede viária, pode identificar quais limiares de centralidade têm maior impacto no IDHM, oferecendo insumos mais concretos para políticas de infraestrutura. Em segundo lugar, a investigação da influência do Nordeste nos modelos pode esclarecer em que medida o baixo desenvolvimento dos municípios da região é explicável por fatores de sofisticação produtiva versus por determinantes históricos e estruturais de outra natureza.


\section*{Apêndice A: Resultados Detalhados do \textit{Grid Search}}

As tabelas a seguir registram os resultados completos da busca em grade para as Árvores de Decisão, fundamentando a escolha dos hiperparâmetros discutidos no texto.

\begin{table}[htbp]
\centering
\caption{Resultados da Busca em Grade para Árvores de Decisão - Municípios}
\label{tab:grid_municipios}
\begin{tabular}{llccrr}
\toprule
\textbf{Variante} & \textbf{max\_depth} & \textbf{min\_leaf} & \textbf{MAE} & \textbf{R²} \\
\midrule
Com Rede & 7 & 10 & 0.0270 & 0.7647 \\
Com Rede & 7 & 2  & 0.0272 & 0.7613 \\
Com Rede & 7 & 5  & 0.0272 & 0.7609 \\
Com Rede & 5 & 10 & 0.0278 & 0.7538 \\
Com Rede & 5 & 2  & 0.0279 & 0.7537 \\
Com Rede & 5 & 5  & 0.0279 & 0.7533 \\
Sem Rede & 5 & 2  & 0.0294 & 0.7277 \\
Sem Rede & 5 & 5  & 0.0294 & 0.7285 \\
Sem Rede & 5 & 10 & 0.0294 & 0.7286 \\
Sem Rede & 7 & 10 & 0.0297 & 0.7219 \\
Sem Rede & 7 & 5  & 0.0298 & 0.7193 \\
Sem Rede & 7 & 2  & 0.0299 & 0.7151 \\
\bottomrule
\end{tabular}
\end{table}

\begin{table}[htbp]
\centering
\caption{Resultados da Busca em Grade para Árvores de Decisão - Regiões Imediatas}
\label{tab:grid_regioes_imediatas}
\begin{tabular}{llccrr}
\toprule
\textbf{Variante} & \textbf{max\_depth} & \textbf{min\_leaf} & \textbf{MAE} & \textbf{R²} \\
\midrule
Com Rede & 5 & 10 & 0.0247 & 0.7788 \\
Com Rede & 5 & 5  & 0.0250 & 0.7725 \\
Com Rede & 5 & 2  & 0.0253 & 0.7698 \\
Com Rede & 7 & 10 & 0.0253 & 0.7686 \\
Com Rede & 7 & 5  & 0.0262 & 0.7508 \\
Com Rede & 7 & 2  & 0.0272 & 0.7229 \\
Sem Rede & 7 & 10 & 0.0254 & 0.7646 \\
Sem Rede & 5 & 10 & 0.0255 & 0.7661 \\
Sem Rede & 5 & 5  & 0.0256 & 0.7614 \\
Sem Rede & 5 & 2  & 0.0261 & 0.7579 \\
Sem Rede & 7 & 5  & 0.0265 & 0.7430 \\
Sem Rede & 7 & 2  & 0.0284 & 0.6996 \\
\bottomrule
\end{tabular}
\end{table}

\clearpage
\putbib[referencias_pt]
\end{bibunit}

\end{document}